\documentclass[a4paper,12pt]{article}

\usepackage[hmargin=3.5cm,vmargin=3cm]{geometry}
\usepackage{calc}
\usepackage{xspace}

\usepackage[komastyle,ilines]{scrpage2}
\setheadtopline{0pt} 
\setheadsepline{1pt}
\ohead{A two-species hydrodynamic model}
\ihead{Laurent Navoret}

 \ifoot{}
 \ofoot{}

\usepackage{fontenc}[T1]
\usepackage[utf8x]{inputenc} 
\usepackage{amsfonts}	
\usepackage{amsmath} 	
\usepackage{amssymb}
\usepackage{amsthm}
\usepackage{array}
\usepackage{psfrag}
\usepackage{subfigure}

\usepackage{enumerate} 	

\usepackage{graphicx} 	
\usepackage{placeins}	

\usepackage[dvips]{hyperref}
\hypersetup{linktocpage=true,colorlinks=true,linkcolor=BlueViolet,citecolor=BlueViolet}

\usepackage[dvipsname,svgnames]{xcolor}

\usepackage{booktabs}

\newcommand{\R}{\mathbb{R}}

\renewcommand{\S}{\mathbb{S}} 

\newcommand{\bomega}{\boldsymbol{\omega}}
\newcommand{\bOmega}{\boldsymbol{\Omega}}
\newcommand{\bj}{\boldsymbol{j}}
\newcommand{\bJ}{\boldsymbol{J}}
\newcommand{\bX}{\boldsymbol{X}}
\newcommand{\bY}{\boldsymbol{Y}}
\newcommand{\bx}{\boldsymbol{x}}
\newcommand{\by}{\boldsymbol{y}}
\newcommand{\bv}{\boldsymbol{v}}
\newcommand{\bV}{\boldsymbol{V}}
\newcommand{\bS}{\boldsymbol{S}}

\newcommand{\eps}{\varepsilon}

\theoremstyle{plain}
\newtheorem{thm}{Theorem}{}
\newtheorem{lem}[thm]{Lemma}
\newtheorem{prop}[thm]{Proposition}

\theoremstyle{definition}

\theoremstyle{remark}

\title{A two-species hydrodynamic model of particles interacting through
self-alignment}
\author{Laurent \textsc{Navoret}\footnote{Institut de Recherche Math\'ematique Avanc\'ee, UMR 7501, Universit{\'e} de Strasbourg et CNRS, 7 rue Ren\'e Descartes, 67000
Strasbourg, France (laurent.navoret@math.unistra.fr).}}
\date{}

\begin{document}

\maketitle

\paragraph{Abstract} In this paper, we present a two-species Vicsek model, that describes alignment interactions of self-propelled particles which can either move or not. The model consists in two populations with distinct Vicsek dynamics \cite{95_Vicsek_PhasTrans2d} that interact only via the passage of the particles from one population to the other. The derivation of a macroscopic description of this model is performed using the methodology used  in \cite{2008_ContinuumLimit_DM} for the Vicsek model: we find out a regime where alignment in the whole population occurs. We obtain a new  macroscopic model for the densities of each populations and the common mean direction of the particles. The treatment of the non-conservativity of the interactions requires a detail study of the linearised interaction operator.

\paragraph{Keywords:} Individual based model; sheep behaviour; Vicsek model; asymptotic analysis; orientation interaction; hydrodynamic limit; collision invariants

\section{Introduction}

The modelling of flocking behaviour, like flock of birds or school of fish, has recently been the subject of a vast literature: one of the main issue tackled in these works is the emergence of collective movement from only local interactions between neighbouring animals, without leaders.  To describe such dynamics, a first modeling approach is to consider a system of self-propelled particles, i.e. particles moving with a preferred speed, and to propose interaction rules: a first class of models have considered binary attraction-repulsion interactions \cite{2009_Carillo_DbleMilling,2007_DOrsognaBertozzi}. These models can be completed by alignment interactions \cite{1982_Aoki,2002_Couzin}. Note that empirical studies on flock of starlings \cite{2008_Ballerini,2010_Cavagna_EmpData} have also been carried out to characterise the interaction law. A vast literature also focuses only on alignment interactions dynamics (without attraction-repulsion interactions) and in particular on the Vicsek model \cite{95_Vicsek_PhasTrans2d}:  each particles tends to align with the mean direction of its neighbours.  In this paper, we propose a model where the particles also follow Vicsek alignment interactions  but can be either moving or at rest. This model is motivated by the displacement dynamics of gregarious animals, like the movement of sheep herds.  The population is thus split in two phases with two independent alignment dynamics: one phase made of particles with a prescribed non-zero speed and the other phase made of zero-speed particles. The passage of the particles from the moving to the non-moving state (and inversely) makes the two phases interact and so raises the question whether the large scale alignment in each phases and the alignment of the two phases themselves could occur. 

To describe and to characterise the large scale dynamics of animal populations, we are interested in macroscopic models that provide the dynamics of macroscopic quantities like the density or the mean velocity inside the flock. Some of these models are phenomenological (e.g. \cite{1999_NonLocalModel_MogKesh,06_Nonlocal_TopazBertozzi}) while others are obtained as the mean-field limits of individual based models \cite{2009_Carillo_DbleMilling,2011_DegondMotsch_PTWMacro,2007_DOrsognaBertozzi,2010_CarilloKlar}. Concerning the Vicsek model, several macroscopic (kinetic or continuum) models have been proposed \cite{2006_BoltzSelfPropel_BertinGregoire,2000_CzirokVicsek} but Degond and Motsch \cite{2008_ContinuumLimit_DM} first provided a mathematical derivation from the microscopic model. Such derivations raise interesting questions about propagation of chaos \cite{2011_BolleyCanCar_Vicsek,2010_BolleyCanCar_StocMFlimit}, and enable to study the convergence rate to the long-time asymptotic flocking states \cite{2007_EmergenceFlocks_CuckerSmale,2009_ProofCuckerSmale_HaLiu,2008_Flocking_HaTadmor,2010_CarFornRosTosc_AsymptFlock}. As the original Vicsek algorithm \cite{95_Vicsek_PhasTrans2d,2004_GregChate_Onset,2003_AldanaHuepe,2004_GregChate_Onset,2011_FrouvLiu_PhaseTrans}, macroscopic models can also present phase transition phenomena \cite{2011_FrouvLiu_PhaseTrans} from disordered to ordered configurations. Several variants of the Vicsek model have been studied: the noise can be implemented in different ways \cite{2008_ChateGin_CollMot}, the noise and the alignment frequency can be made dependent on the local density, an angle of vision can be added \cite{2010_Frouvelle_Vic}. We also refer to \cite{2010_CarilloForn_Review} for a review on swarming models. We mention that models have also been proposed for vehicular/pedestrian traffic \cite{2001_Helbing_RevTraffic,2011_BellomoDogbe_traffic} but they differ from the previous ones since each particle then pursue a goal.

As announced above, we study here a variant of the Vicsek model for the displacement of a system of particles which can either move or not. Indeed, this work is motivated by the behaviour of sheep in groups during grazing period: sheep alternate between motionless time when grazing and displacements to look for fresher grass. In our model, the (biological) state of each particles is then described by a speed variable which takes only two values. The change between the two states of the animal can be either spontaneous or triggered by fellows. Moreover, as biologists experimentally show it in \cite{2009_Pillot}, any member of the group can initiate a movement: this is called distributed leadership. Therefore, in the model we consider, the change of speed is modelled by a Markovian jump process, whose jump rates are depending on the local alignment to the neighbouring particles:  the more one motionless particle is aligned with moving particles, the more it is likely to interact and then to change its speed. Comparisons with experimental data should be required to confirm or improve such interacting law. 
For more information about herding, we refer to \cite{2010_Pillot_CollMovInitStop,2011_Pillot_Scalable,2003_Vertebrates_CouzinKrause}. 
Here after, we will study this model as a minimal model of synchronization between two phases with different characteristic speeds and exchanges of individuals between the two phases.

Based on this microscopic dynamics, a mean field kinetic equation with a discrete speed dynamics is considered: the Vicsek model is coupled with a Markovian jump process modelling the change of speed. This kinetic model is the starting point of our study. This model is then discrete in the speed variable and thus shares some analogies with the discrete kinetic models developed for traffic modeling in \cite{2007_DelitalaTosin_DiscreteKinetic,2011_BellomoDogbe_traffic}. The discreteness  of the velocity modulus space is not a mathematical simplification (as it is the case for the Broadwell or Carleman models  \cite{1975_Gatignol}), but is here a modelling assumption: as in traffic dynamics, the distribution function in the speed variable is really discontinuous \cite{2009_Pillot}.  However, we note that our model strongly differs from all these works since the velocity direction space is kept here as a continuum, i.e. the unit circle $\S^1$. Note that Markovian jump models have also been proposed for the population dynamics of cells, in particular to model proliferating or destructive interactions \cite{2005_BellouDellitala, 2010_Bellomo_Multiscale}: unlike these models, the particles are here conserved in time.

The main purpose of the paper is to investigate the large scale dynamics of this two-phase kinetic model. Two time scales corresponding to the two types of interactions, are involved in this model: the time scale $\varepsilon$ of the Vicsek interactions and the time scale  $\delta$ of the speed changes. Here, we focus on the asymptotic regime $\varepsilon \ll \delta \ll 1$, where the Vicsek interactions are more frequent than the speed changes and where both time scales are small compared to the macroscopic one: to this aim, we first consider the dynamics in the regime $\varepsilon \ll 1, \delta = O(1)$ before investigating the $\delta \rightarrow 0$ limit. Following the methodology introduced in \cite{2008_ContinuumLimit_DM}, in the large scale Vicsek limit ($\eps \rightarrow 0$), we obtain two macroscopic Vicsek models, giving the dynamics of the densities and the mean directions for each phases (moving and non-moving phase), coupled through the speed change operator. The specificity of this model is that the macroscopic speed change operator do not conserve the global mean direction. Indeed, they are derived from the integration of the kinetic speed change operator against ``generalised collisional invariants'' \cite{2008_ContinuumLimit_DM} which are specific for each Vicsek operators: they depend on the parameters of the Vicsek operator, which can be different for the moving and the non-moving phase. 

The second step now consists in obtaining an averaged two-phase model, once the equilibria of the speed change dynamics  are reached. The derivation of such ``simplified'' averaged models is a very challenging issue in two-phase fluid dynamics and has been the subject of a lot of works \cite{1975_Ishii}. To achieve our goal, we first find out the equilibria of the speed-change operator: thanks to a careful study of the speed change operator, we are able to show that the two phases are either locally aligned or locally in the opposite direction. Moreover, the densities of the two phases are linked through a non-linear balance equation. To find out the dynamics of these equilibria, we face once again the non-conservativity nature of the model: indeed, this property prevent us from having obvious ``collisional invariants'' that would result in balance simplifications. To overcome this difficulty, we perform an Hilbert expansion around the equilibria and we figure out the kernel and the image of the linearised exchange operator, which acts on the two-dimensional space of the mean directions: any non-zero element of the one-dimensional orthogonal space of its image then provide a ``generalised collisional invariants''. Supposing that the equilibria are reached, we thus find out the dynamical system satisfied by the total density and the common direction of the two-phases.

 All this methodology enables to provide the dynamics of the two-phases, once they aligned. It provides a new non-conservative model for swarming population, whose several mathematical properties are yet left open. For instance, the stability of the equilibria and the hyperbolicity of the macroscopic model will be investigated in future works. In addition, this model raises some interesting questions about the modelling of alignment interactions in herds: what is the macroscopic behaviour in other regime of parameters (for instance if the Vicsek interactions are less frequent than the speed changes) ? We can ask also about the appearance of phase transitions in such discrete speed dynamics.

 The outline of this article is as follows. In Section \ref{Sec:Vicsek2phase_micro_kinetic}, we introduce the two-phase model at the particle and the kinetic level. In Section \ref{Sec:macro_dyn}, we present the macroscopic regime we are interested in: we perform a hydrodynamic rescaling and uncouple the Vicsek and the speed change time scales. We then state the two main results. Sections \ref{Sec:LargeScale_Vicsek} and \ref{Sec:LargeScale_speedchange} are devoted to the proofs of the derived hydrodynamic models: as explained above, we first focus on the large scale Vicsek dynamics (sect. \ref{Sec:LargeScale_Vicsek}) and then in the large scale speed change asymptotic (sec. \ref{Sec:LargeScale_speedchange}). In the two steps, we figure out the equilibria and close the system using collisional invariants. Appendices \ref{annex:momentum_computations}, \ref{Appendix:momentum_balance} and \ref{Appendix:CollisionalInvariant} provide some detailed computations.

\section{A two-speed Vicsek model}
\label{Sec:Vicsek2phase_micro_kinetic}

We present in this section an individual-based model and its mean-field kinetic version to describe alignment interactions in a system of self-propelled particles, that can move and stop.  

\subsection{The microscopic model}
\label{Sec:micro_model}
We consider $N$ particles with positions $\bX_{k} \in \R^2$ and velocities $\bV_{k} = c\eta_{k}\bomega_{k} \in \R^2$ for $k \in \left\{1,\ldots ,N\right\}$, where $\eta_{k} \in \left\{0,1\right\}$ and $\bomega_{k} \in \S^{1} = \left\{\bomega \in \R^2, |\bomega| = 1\right\}$ denote respectively the velocity moduli and the velocity directions. The magnitude of the velocities can take only two values $0$ or $c > 0$, and then the particles are separated into two subgroups: the subgroup $\left\{k,\ |\bV_{k}| = 0\right\}$ made of the particles at rest and the subgroup $\left\{k,\ |\bV_{k}| = c\right\}$ of the moving particles. 

\paragraph{The Vicsek dynamics within the subgroups.} The interactions among particles of the same phase are given by the Vicsek model, as described in \cite{2008_ContinuumLimit_DM}: a particle is supposed to have a mimetic behaviour with the neighbouring congeners being in the same state (moving or at rest). The dynamics of positions $\bX_{k}$ and orientations $\bomega_k$ are given by the following equations:
\begin{equation}
\begin{split}
&\frac{d\bX_{k}}{dt} = c\,\eta_{k}\bomega_{k},\\
&d\bomega_{k} = (\text{\bf Id} - \bomega_{k}\otimes\bomega_{k})(\nu_{k}\bar{\bomega}_{k}dt + \sqrt{2d_{k}}d\bold B_{t}),
\end{split}
\label{Eq:Vicsek_part}
\end{equation}
where $\text{\bf Id}$ denotes the identity matrix, $\bold w\otimes\bold v$ denotes the tensor product of the two vectors $\bold w$ and $\bold v$. The operator $(\text{\bf Id} - \bomega_{k}\otimes\bomega_{k})$ is the projection operator onto the orthogonal plane to $\bomega_{k}$: it ensures the norm of the direction $\bomega_{k}$ to be unity. Two dynamics are in competition: each particle tend to align with the mean direction $\bar{\bomega}_{k}$ of their neighbours in the disc of radius $R$ around them :
\begin{equation}
\bar{\bomega}_{k} = \frac{\bold J_{k}}{|\bold J_{k}|},\quad \bold J_{k} = \displaystyle\sum_{\begin{subarray}{c}
j,\ \eta_j = \eta_k,\\
|\bX_{j} -\bX_{k}| \leq R
\end{subarray}} \bomega_{j},
\end{equation}
and noise is applied to the direction with a Brownian motion $\bold B_{t}$ on $\R^2$. Note that in the definition of the local mean direction $\bold J_{k}$, only the particles of the same phase are taken into account. This two behaviours are quantified by the alignment intensities $\nu_k$ and the noise intensities $d_k$. These two parameters are uniform in each subgroup:
\begin{equation*}
(\nu_k,d_k) = \begin{cases}
(\nu_0,d_0),&\text{ if } \eta_k = 0,\\
(\nu_1,d_1),&\text{ if } \eta_k = 1.
\end{cases}
\end{equation*}
At this level, the dynamics of the two phases are totally independent: the moving particles follows the Vicsek model with parameters $(\nu_1,d_1)$ and the motionless particles\footnote{Note that particles at rest are not moving in space but their directions are changing in time.} follows a static Vicsek model with parameter $(\nu_0,d_0)$. The static Vicsek model is similar to the Ising model, that provides the dynamics of spins distributed on a lattice.

\paragraph{The speed change: the Markov process $\eta_k$.} The particles can also change their speeds  $\eta_k$ from the moving ($\eta_k = 1$) to the motionless state ($\eta_k = 0$) and from the motionless to the moving state: it results in a permanent exchange of the particles between the moving and the unmoving phases. This exchange between the two subgroups are described by the dynamics of $\eta_{k}$: it is a time-continuous Markov process on the state space $\left\{0,1\right\}$. The transition rates are given by:
\begin{equation}
g_{k} = \tau_{k}\left[1+\alpha\frac{1}{N}\sum_{\begin{subarray}{c}
j,\ \eta_{j} \neq \eta_{k},\\ 
|\bX_{j} - \bX_{k}| \leq R
\end{subarray}}\frac{(1+\bomega_{k}\cdot\bomega_{j})}{2}\right],
\label{Eq:rate}
\end{equation}
where $\tau_k$ is the intrinsic rate which can be can take different values for the two subgroups:
\begin{equation*}
\tau_k = \begin{cases}
\tau_0,&\text{ if } \eta_k = 0,\\
\tau_1,&\text{ if } \eta_k = 1.
\end{cases}
\end{equation*}
The second term in the sum is of order $\alpha\tau_k$ and makes the rate depend on the local alignment with the members of the other phase :  a particle at rest (resp. moving) is all the more likely to change its speed as it is locally aligned with its neighbouring moving congeners (resp. congeners at rest). It is aimed at describing the distributed leadership observed in herds \cite{2009_Pillot}: each particle can bring forth a speed change of its neighbours that are aligned with it. 
 
The two Vicsek models are now coupled via the passage of the particles from one subgroup to the other. The dependence of the transition rates on the alignment of the two subgroups might bring forth the alignment of the whole population. The goal of this paper is to determine for which set of parameters, alignment in the whole system occurs.

\subsection{The mean field kinetic model}

We introduce the two distribution functions in phase space: $f_{0}(\bx,\bomega,t)$ for the particles at rest and $f_{1}(\bx,\bomega,t)$ for the moving particles. The mean field model we consider is the following:
\begin{align}
\partial_{t} f_{0} &=  \mathcal{Q}_{0}(f_{0}) + \mathcal{E}(f_{0},f_{1}),\label{Eq:f0}\\
\partial_{t} f_{1} + c\,\bomega\cdot\nabla_{\bx}f_{1} &=  \mathcal{Q}_{1}(f_{1}) - \mathcal{E}(f_{0},f_{1}),\label{Eq:f1}
\end{align}
where $\nabla_{\bx}$ denotes the space gradient operator. The left-hand sides of these two equations are the transport operators of the particles with velocities $0\times\bomega$ and $1\times \bomega$, while the right-hand sides model the velocity dynamics of the particles. 

The operator $\mathcal{Q}_{0}$ and $\mathcal{Q}_{1}$ are the Vicsek operators:
\begin{align}
&\mathcal{Q}_0(f_0) = - \nabla_{\bomega}\cdot(\nu_{0}\mathcal{F}[f_{0}]f_{0}) + d_{0}\Delta_{\bomega}f_{0},\\
&\mathcal{Q}_1(f_1) = - \nabla_{\bomega}\cdot(\nu_{1}\mathcal{F}[f_{1}]f_{1}) + d_{1}\Delta_{\bomega}f_{1},
\end{align}
with $\nabla_{\bomega}\cdot$ and $\Delta_{\bomega}$ are respectively the divergence and the Laplace operators\footnote{If $\theta$ is the polar coordinate associated to an orthonormal basis $(\bold e_1,\bold e_2)$ of $\R^2$, then the divergence of a scalar function $f(\bomega)$ is given by $\partial_\theta f$ and the divergence of a tangent vector field $\bold A = A_\theta \bold e_\theta$, where $\bold e_\theta = (-\sin\theta,\cos\theta)$ is the local polar basis, is given by $\partial^2_{\theta^2} A_\theta$.} on the circle $\S^1$. Note that the operators, $\mathcal{Q}_{0}$ and $\mathcal{Q}_{1}$, differ from one to each other only in the couple of parameters $(\nu_0,d_0)$ and $(\nu_1,d_1)$. The operator $\mathcal{F}$ denotes the alignment forces in each group and is given by:
\begin{align}
&\mathcal{F}[f](\bx,\bomega ,t) = (\text{\bf Id} - \bomega\otimes\bomega)\bar{\bomega}[f](\bx ,t),\\
&\bar{\bomega}[f](\bx ,t) = \frac{\mathcal{J}[f](\bx,t)}{|\mathcal{J}[f](\bx,t)|},\ \mathcal{J}[f](\bx,t) = \int_{\by\in \R^{2}, \bv\in \S^{1}} K(|\bx-\by|)\bv f(\by,\bv,t) d\by d\bv,
\end{align}
where $K(|\bx|)$ is the interaction kernel equal to the indicator function of the disc of radius $R$. These operators was derived in \cite{2008_ContinuumLimit_DM}: without Brownian motion ($d_0 = d_1 = 0$) nor the operator $\mathcal{E}(f_{0},f_{1})$, equations (\ref{Eq:f0})-(\ref{Eq:f1}) are also satisfied by the empirical distribution functions of the particles following the Vicsek rules \eqref{Eq:Vicsek_part} and so the mean-field limit can be investigated: a rigorous study, carried out in \cite{2011_BolleyCanCar_Vicsek}, justifies the resulting kinetic equations for the one-particle distribution functions. We refer also to \cite{spohn_book} for theoretical developments on mean-field limits. Moreover, numerical simulations \cite{2010_MotschLN_NumMacroVic} provide a numerical validation of this kinetic description at least in some range of parameters. 

The exchange term $\mathcal{E}(f_{0},f_{1})$ in the right-hand sides of both (\ref{Eq:f0}) and (\ref{Eq:f1}) is given by:
\begin{align}
&\mathcal{E}(f_{0},f_{1}) = - \tau_{0}\mathcal{G}[f_{1}]f_{0} + \tau_{1}\mathcal{G}[f_{0}]f_{1},\label{Eq:E_firstdef}\\
&\mathcal{G}[f](\bx,\bomega,t) = 1 + \alpha\int_{\by\in \R^{2}, \bv\in \S^{1}}K(|\bx-\by|)\frac{(1+\bv\cdot\bomega)}{2} f(\by,\bv,t)d\by d\bv.\label{Eq:G_firstdef}
\end{align}
Without the Vicsek operators, equations (\ref{Eq:f0})-(\ref{Eq:f1}) formally represent the forward Kolmogorov equation of the time-inhomogeneous Markov process defined by the rates \eqref{Eq:rate}. We refer to \cite{2005_Stroock_MarkovProcess} for more detail on this subject. 

System (\ref{Eq:f0}-\ref{Eq:G_firstdef}) is the starting point of our study of the large time and space scale dynamics of the microscopic model presented in section \ref{Sec:micro_model}.

\section{The macroscopic dynamics: rescaling and main results}
\label{Sec:macro_dyn}

In this section, we focus on the large scale dynamics of system (\ref{Eq:f0}-\ref{Eq:G_firstdef}). We are specifically interested in regimes where the time scale of the two interactions present in the model, the internal Vicsek dynamics inside the subgroups and the exchange dynamics between the subgroups, are small compared with the time scale of observation. To investigate asymptotic regimes, we first begin by performing a time and space rescaling to obtain a dimensionless system.

Let us write system (\ref{Eq:f0}-\ref{Eq:G_firstdef}) is rescaled variables. Let $\nu_V$ the typical Vicsek interaction frequency, $\nu_0 = \nu_V\nu_0'$, $\nu_1 = \nu_V\nu_1'$, and $\tau_E$ the typical speed change frequency, $\tau_0 = \tau_E\tau_0'$, $\tau_1 = \tau_E\tau_1'$. The dimensionless diffusion coefficients are also given by: $d_0 = \nu_V d_0'$, $d_1 = \nu_V d_1'$. We then introduce the dimensionless time and space variables: $t' = \nu_V t$, $x' = x\nu_V/c$. After dropping the tildes, system (\ref{Eq:f0}-\ref{Eq:G_firstdef}) becomes in the new variables:
\begin{align}
\partial_{t} f_{0} &=  \mathcal{Q}_{0}(f_{0}) + \frac{1}{\delta}\,\mathcal{E}(f_{0},f_{1}),\\
\partial_{t} f_{1} + \bomega\cdot\nabla_{\bx}f_{1} &=  \mathcal{Q}_{1}(f_{1}) - \frac{1}{\delta}\,\mathcal{E}(f_{0},f_{1}),
\end{align}
where $\delta = \nu_V/\tau_E$ quantifies the relative intensity of the two interactions and is supposed to be order $O(1)$. All the coefficients of the operators $\mathcal{Q}_{0}$, $\mathcal{Q}_{1}$ and  $\mathcal{E}$ (i.e. $\nu_0,\ \nu_1,\ d_0,\ d_1,\ \tau_0,\ \tau_1)$ are now dimensionless and are supposed to be of order $O(1)$.

\subsection{Hydrodynamic rescaling} 
We here perform a hydrodynamic rescaling to look at the large time and space scale dynamics. The hydrodynamic rescaling consists in introducing macroscopic variables in space and time: $\tilde{\bx} = \eps \bx$, $\tilde{t} = \eps t$, with $\eps \ll 1$. In the new variables, the distribution function of the zero speed particles $f_0^{\eps}(\tilde{\bx},\bomega,\tilde{t}) = f_0(\bx,\bomega,t)$ and of the moving particles $f_1^{\eps}(\tilde{\bx},\bomega,\tilde{t}) = f_1(\bx,\bomega,t)$ satisfy the following system (dropping the tildes):
\begin{align}
\eps(\partial_{t} f_{0}^{\eps}) &= \mathcal{Q}_{0}^{\eps}(f_{0}^{\eps}) + \frac{1}{\delta}\,\mathcal{E}^{\eps}(f_{0}^{\eps},f_{1}^{\eps}),\label{Eq:f0_eps}\\
\eps(\partial_{t} f_{1}^{\eps} + \bomega\cdot\nabla_{\bx}f_{1}^{\eps}) &= \mathcal{Q}_{1}^{\eps}(f_{1}^{\eps}) - \frac{1}{\delta}\,\mathcal{E}^{\eps}(f_{0}^{\eps},f_{1}^{\eps}).\label{Eq:f1_eps}
\end{align}
The Vicsek operators $\mathcal{Q}_{0}^{\eps}$ and $\mathcal{Q}_{1}^{\eps}$ and the exchange operator $\mathcal{E}^{\eps}$ are given by:
\begin{align}
&\mathcal{Q}_{0}^{\eps}(f_{0}^{\eps}) = - \nabla_{\bomega}\cdot(\nu_{0}\mathcal{F}^{\eps}[f_{0}^{\eps}]f_{0}^{\eps}) + d_{0}\Delta_{\bomega}f_{0}^{\eps},\\
&\mathcal{Q}_{1}^{\eps}(f_{1}^{\eps}) = - \nabla_{\bomega}\cdot(\nu_{1}\mathcal{F}^{\eps}[f_{1}^{\eps}]f_{1}^{\eps}) + d_{1}\Delta_{\bomega}f_{1}^{\eps},\\
&\mathcal{E}^{\eps}(f_{0}^{\eps},f_{1}^{\eps}) = - \tau_{0}\mathcal{G}^{\eps}[f_{1}^{\eps}]f_{0}^{\eps} + \tau_{1}\mathcal{G}^{\eps}[f_{0}^{\eps}]f_{1}^{\eps},
\end{align}
where $\mathcal{F}^{\eps}$ is the rescaled interaction forces given by:
\begin{align}
&\mathcal{F}^{\eps}[f^{\eps}](\bx,\bomega ,t) = (\text{\bf Id} - \bomega\otimes\bomega)\bar{\bomega}^{\eps}[f^{\eps}](\bx ,t),\nonumber\\
&\bar{\bomega}^{\eps}[f^{\eps}](\bx ,t) = \frac{\bJ^{\eps}[f^{\eps}](\bx,t)}{|\bJ^{\eps}[f^{\eps}](\bx,t)|},\ \bJ^{\eps}[f^{\eps}](\bx,t) = \frac{1}{\eps^{2}}\int_{\by\in \R^{2},\bv\in \S^{1}} K\left(\left|\frac{\bx -\by}{\eps}\right|\right)\bv f^{\eps}(\by ,\bv ,t) d\by d\bv,\\
\intertext{and $\mathcal{G}^{\eps}$ is the rescaled coupling coefficient:}
&\mathcal{G}^{\eps}[f^{\eps}](\bx,\bomega,t) = 1 + \frac{\alpha}{\eps^{2}}\int_{\by\in \R^{2}, \bv\in \S^{1}}K\left(\left|\frac{\bx -\by}{\eps}\right|\right)\frac{(1+\bv\cdot\bomega)}{2} f^{\eps}(\by ,\bv ,t)d\by d\bv.\label{Eq:g_eps}
\end{align}
Let us make an expansion with respect to $\eps$ of these interaction terms. To this aim, let us introduce the macroscopic quantities related to a distribution function $f(\bx,\bomega,t)$, that is the density $\rho[f](\bx,t)$, the momentum $\bj[f](\bx,t)$ and the mean direction $\bOmega[f](\bx,t)$: 
\begin{equation*}
\begin{split}
&\rho[f](\bx ,t) = \int_{\bv\in \S^{1}} f(\bx ,\bv, t) dv,\\
&\bj[f](\bx ,t) = \int_{\bv\in \S^{1}} \bv f(\bx ,\bv, t) d\bv,\quad \bOmega[f](\bx,t) = \frac{\bj[f](\bx ,t)}{\left|\bj[f](\bx ,t)\right|}.
\end{split}
\end{equation*}
The following lemma provides the expansion of $\bar{\bomega}^{\eps}$ and $\mathcal{G}^{\eps}$ with respect to $\eps$. 
\begin{lem} We have the following expansions: 
\begin{align}
&\bar{\bomega}^{\eps}(\bx ,t) = \bOmega[f^{\eps}](\bx ,t) + O(\eps^{2}),\\
&\mathcal{G}^{\eps}[f^{\eps}](\bx,\bomega,t) = 1 + \alpha\frac{\rho[f^{\eps}](\bx ,t) + \bomega\cdot \bj[f^{\eps}](\bx ,t)}{2} + O(\eps^{2}),
\end{align}
where $\rho[f^{\eps}]$, $\bj[f^{\eps}]$, $\bOmega[f^{\eps}]$ are the density, the momentum and the mean direction related to $f^{\eps}$.
\label{Lemma:Vicsek_expansion}
\end{lem}
\noindent The proof of Lemma~\ref{Lemma:Vicsek_expansion} is omitted. Lemma~\ref{Lemma:Vicsek_expansion} enables to write system (\ref{Eq:f0_eps}-\ref{Eq:g_eps}) as following:
\begin{align}
\partial_{t} f_{0}^{\eps} &= \frac{1}{\eps}Q_{0}(f_{0}^{\eps}) + \frac{1}{\eps\delta}E(f_{0}^{\eps},f_{1}^{\eps}) + O\left(\eps\right),\label{Eq:f0_eps2}\\
\partial_{t} f_{1}^{\eps} + \bomega\cdot\nabla_{\bx}f_{1}^{\eps} &= \frac{1}{\eps}Q_{1}(f_{1}^{\eps}) - \frac{1}{\eps\delta}E(f_{0}^{\eps},f_{1}^{\eps}) + O\left(\eps\right),\label{Eq:f1_eps2}
\end{align}
where $Q_{0}$, $Q_{1}$ and $E$ are the operators of order $O(1)$ in the expansions of $\mathcal{Q}_{0}^\eps$, $\mathcal{Q}_{1}^\eps$ and $\mathcal{E}^\eps$. They are given by:
\begin{align}
&Q_{0,1}(f_{0,1}^{\eps}) = - \nabla_{\bomega}\cdot(\nu_{0,1}F[f_{0,1}^{\eps}]f_{0,1}^{\eps}) + d_{0,1}\Delta_{\bomega}f_{0,1}^{\eps},\label{Eq:Vicsek_kernel}\\
&E(f_{0}^{\eps},f_{1}^{\eps}) = - \tau_{0}g[f_{1}^{\eps}]f_{0}^{\eps} + \tau_{1}g[f_{0}^{\eps}]f_{1}^{\eps},\label{Eq:Exchange_op}
\intertext{where the expressions of $F$ and $g$ result from lemma~\ref{Lemma:Vicsek_expansion}:}
&F[f](\bx ,\bomega ,t) = (\text{\bf Id} - \bomega\otimes\bomega)\bOmega[f](\bx ,t),\\
&g[f](\bx ,\bomega ,t) = 1 + \alpha\frac{\rho[f](\bx ,t) + \bomega\cdot \bj[f](\bx ,t)}{2}.\label{Eq:g_eps2}
\end{align}
Note that equation \eqref{Eq:Vicsek_kernel} contains terms indexed by ``$0,1$'' : in all the following, an equation whose all the quantities $f_{a,b}$ are indexed by ``$a,b$'' represents the set of two equations\footnote{For instance, $f_{0,1} = k_{1,0}$ means $f_{0} = k_{1}$ and $f_{1} = k_{0}$.}, the equation with the left indexes ``$a$'' and the equation with the right indexes ``$b$''.

\subsection{The macroscopic regime: main results}

We would like now to investigate the hydrodynamic limit $\eps \rightarrow 0$ in system (\ref{Eq:f0_eps2}-\ref{Eq:g_eps2}). It corresponds to a regime where the Vicsek interaction and the speed change time scales are both small compared to the observation time scale and of order $O(\eps)$. However, the different mathematical natures of the Vicsek operators, $Q_{0}$ and $Q_{1}$, and of the exchange operator $E$ prevent us from achieving this goal in one unique step.

Instead of supposing that the Vicsek and exchange time scales are both of  order $O(\eps)$, we begin by supposing that only the Vicsek time scale is so, while the exchange time scale remains of order $O(1)$ at large scale:
\begin{equation*}
\delta' = \eps \delta = O(1).
\end{equation*}
%
We here uncouple the time scales of the two interactions and first link the macroscopic time scales to the Vicsek interactions only. In the following, we will provide the asymptotic dynamics of system (where we omit the prime):
\begin{align}
\partial_{t} f_{0}^{\eps} &= \frac{1}{\eps}Q_{0}(f_{0}^{\eps}) + \frac{1}{\delta}E(f_{0}^{\eps},f_{1}^{\eps}) + O\left(\eps\right),\label{Eq:f0_eps_delta}\\
\partial_{t} f_{1}^{\eps} + \bomega\cdot\nabla_{\bx}f_{1}^{\eps} &= \frac{1}{\eps}Q_{1}(f_{1}^{\eps}) - \frac{1}{\delta}E(f_{0}^{\eps},f_{1}^{\eps}) + O\left(\eps\right),\label{Eq:f1_eps_delta}
\end{align}
in the limit:
\begin{equation}
\eps \rightarrow 0,\quad \delta = O(1).\label{Limit1}
\end{equation}
In a second step, we will take the limit: 
\begin{equation}
\delta \rightarrow 0,\label{Limit2}
\end{equation}
It consists in considering that the time scales of the exchange interactions are also small compared to the time scale of the observation. Note that instead of first taking a small Vicsek interaction time scale and then a small exchange time scale, we would make it in the opposite order. Such a study would be the subject of future works. 

In the two following theorems, the two limits (\ref{Limit1}) and (\ref{Limit2}) in system (\ref{Eq:f0_eps_delta}-\ref{Eq:f1_eps_delta}) provides the macroscopic dynamics in the regime $\eps \ll \delta \ll 1$. Before stating the results, let us introduce the Von-Mises velocity distributions $M_{\lambda,\bOmega}$:
\begin{equation}
M_{\lambda,\bOmega}(\bomega) = C_{\lambda}\exp\left(\frac{\bomega\cdot\bOmega}{\lambda}\right),\quad
\int_{\bomega\in\S^{1}} M_{\lambda,\bOmega}(\bomega) d\bomega = 1.\label{Eq:vonMises}
\end{equation}
where $\lambda$ stands for the temperature of the distribution and $C_{\lambda}$ are a re-normalisation constant to ensure the total mass to be unity. The following theorem states the hydrodynamic system satisfied by the densities and the mean directions of the two phases in the regime (\ref{Limit1}):
\begin{thm} \textbf{[Limit $\eps \rightarrow 0,\ \delta = O(1)$].} \begin{enumerate}
\item The limits of the distribution functions as $\eps$ goes to $0$ are given by:
\begin{equation*}
f_0^{\eps}(\bx ,\bomega ,t) \rightarrow \rho_{0}(\bx ,t)M_{\lambda_0,\bOmega_0(\bx ,t)}(\bomega),\quad f_1^{\eps}(\bx ,\bomega ,t) \rightarrow \rho_{1}(\bx ,t)M_{\lambda_1,\bOmega_1(\bx ,t)}(\bomega).
\end{equation*}
with $\lambda_{0,1} = d_{0,1}/\nu_{0,1}$.
\item The densities $\rho_{0,1}(\bx,t)$ and the mean directions $\bOmega_{0,1}(\bx,t)$ satisfy the system:
\begin{align}
&\partial_{t} \rho_{0} = \frac{1}{\delta}R(\rho_0,\bOmega_0,\rho_1,\bOmega_1),\label{Eq:rho0_thm_eps}\\
&\partial_{t} \rho_{1} + \nabla_{\bx}\cdot (c_{1}\rho_{1}\bOmega_{1}) = - \frac{1}{\delta}R(\rho_0,\bOmega_0,\rho_1,\bOmega_1),\label{Eq:rho1_thm_eps}\\
&\rho_{0}\partial_{t}\bOmega_{0} = \lambda_{0}(\text{\bf Id}-\bOmega_{0}\otimes\bOmega_{0})\left[-\nabla_{\bx}\rho_{0} + \frac{1}{\delta}\beta_{0}\bS_{0}(\rho_0,\bOmega_0,\rho_1,\bOmega_1)\right],\label{Eq:rhoOmega0_thm_eps}\\
&\rho_{1}\partial_{t}\bOmega_{1} + \gamma_{1}\rho_{1}(\bOmega_{1}\cdot\nabla_{\bx})\bOmega_{1} =\nonumber\\
&\hspace{2cm}\lambda_{1}(\text{\bf Id}-\bOmega_{1}\otimes\bOmega_{1})\left[-\nabla_{\bx}\rho_{1} + \frac{1}{\delta}\beta_{1}\bS_{1}(\rho_0,\bOmega_0,\rho_1,\bOmega_1)\right],\label{Eq:rhoOmega1_thm_eps}
\end{align}
where $c_1$, $\gamma_{1}$, $\beta_{0,1}$ are constants defined in \eqref{Eq:const_c} and \eqref{Eq:const_gamma_beta}. The operators $R$, $\bS_{0}$ and $\bS_{1}$ comes from the exchange dynamics and are defined in \eqref{Eq:exchange_macro} and (\ref{Eq:S_0}-\ref{Eq:S_1}).
\end{enumerate}
\label{thm:limit_eps}
\end{thm}
\noindent Before developing the proof of this theorem in section \ref{Sec:LargeScale_Vicsek}, let us point out the main features of system (\ref{Eq:rho0_thm_eps}-\ref{Eq:rhoOmega1_thm_eps}). First, as the original macroscopic Vicsek model \cite{2008_ContinuumLimit_DM}, system (\ref{Eq:rho0_thm_eps}-\ref{Eq:rhoOmega1_thm_eps}) is non-conservative due to the macroscopic geometric constraints $|\bOmega_{0,1}| =  1$ guaranteed by the projection operators $(\text{\bf Id}-\bOmega_{0,1}\otimes\bOmega_{0,1})$. But, the above two-speed model is all the more non-conservative since the presence of the exchange operators $R$, $\bS_{0,1}$, that model the large scale change of speed of the particles. Note also that, unlike the mass exchange, the momentum exchange operators are not the same for the two subgroups ($\bS_{0} \neq \bS_{1}$) and thus do not compensate.

Then, we take the limit $\delta \rightarrow 0$ in system (\ref{Eq:rho0_thm_eps}-\ref{Eq:rhoOmega1_thm_eps}) to obtain the final result:
\begin{thm} \textbf{[Limit $\delta \rightarrow 0$].} \begin{enumerate}
\item The set of equilibria of the density exchange operator $R$ is given by:
\begin{equation*}
\left\{(\rho_0,\rho_1) \in (\R^+)^2,\ \rho_0 = f_{\Phi}(\rho_1)\right\},
\end{equation*}
where $f_{\Phi}$, defined in \eqref{Eq:density_balance}, is a non-linear function depending on $\bOmega_0$ and $\bOmega_1$ through the local alignment $\Phi$ \eqref{Eq:Local_Align}. 
\item The set of equilibria of the mean direction exchange operators $(\text{\bf Id}-\bOmega_{0}\otimes\bOmega_{0})\bS_{0}$ and $(\text{\bf Id}-\bOmega_{1}\otimes\bOmega_{1})\bS_{1}$ is the following: 
\begin{equation*}
\left\{(\bOmega_0,\bOmega_1) \in (\S^1)^2,\ \bOmega_0 = \bOmega_1\text{ or }\bOmega_0 = -\bOmega_1\right\}.
\end{equation*}
\item Once the set of equilibria $\bOmega_0 = \bOmega_1$ (resp. $\bOmega_0 = -\bOmega_1$) is reached, system (\ref{Eq:rho0_thm_eps}-\ref{Eq:rhoOmega1_thm_eps}) yields  the following closed system for the total density $\rho = \rho_0 + \rho_1$ and the common mean direction $\bOmega = \bOmega_0 = \bOmega_1$ (resp. $\bOmega = \bOmega_1 = -\bOmega_0$):
\begin{align}
&\partial_{t} \rho + \nabla_{\bx}\cdot (c_{1}\rho_1[\rho]\bOmega) = 0,\label{Eq:rho_thm_delta}\\
&M^+(\rho) \partial_{t}\bOmega + \gamma_1 N^+(\rho)\ (\bOmega\cdot\nabla_{\bx})\bOmega = -P^+(\rho)(\text{\bf Id}-\bOmega\otimes\bOmega)\ \nabla_{\bx}\rho, \label{Eq:Omega_thm_delta}\\
\left(resp. \right. &\left. M^-(\rho) \partial_{t}\bOmega + \gamma_1 N^-(\rho)\ (\bOmega\cdot\nabla_{\bx})\bOmega = -P^-(\rho)(\text{\bf Id}-\bOmega\otimes\bOmega)\ \nabla_{\bx}\rho,\right) \label{Eq:Omega_equilibre_opposite}
\end{align}
where $\rho_1[\rho] = (Id + f_\Phi)^{-1}(\rho) $. The functions $M^\pm(\rho)$ and $N^\pm(\rho)$ and $P^\pm(\rho)$ are trilinear with respect to $\rho_1[\rho]$ and $\rho_0[\rho] = \rho - \rho_1[\rho]$. Their explicit expressions are given in proposition~\ref{prop:macro_model}. 
\end{enumerate}
\label{thm:limit_delta}
\end{thm}
\noindent The proof of this theorem is developed in section \ref{Sec:LargeScale_speedchange}. This theorem provides the macroscopic dynamics of the density of particles $\rho$ and the momentum $\Omega$ once the two subgroups of particles are aligned. Note that the derivation of system (\ref{Eq:rho_thm_delta}-\ref{Eq:Omega_thm_delta}) is a priori not obvious since the difference between the momentum exchange operators ($\bS_{0} \neq \bS_{1}$) prevents us from directly cancelling them when summing the two momentum equations (\ref{Eq:rhoOmega0_thm_eps}-\ref{Eq:rhoOmega1_thm_eps}). We overcome this difficulty by carefully analysing the exchange operators and their linearisations. 

\section{The large scale dynamics of the Vicsek interactions:  limit $\eps\rightarrow 0$}
\label{Sec:LargeScale_Vicsek}

This part is devoted to the main steps of the proof of theorem~\ref{thm:limit_eps}, while technical details are put in appendix~\ref{annex:momentum_computations}. The derivation of the macroscopic equations~(\ref{Eq:rho0_thm_eps}-\ref{Eq:rhoOmega1_thm_eps}) from the kinetic system (\ref{Eq:f0_eps_delta}-\ref{Eq:f1_eps_delta}) is similar to the one of the macroscopic Vicsek model \cite{2008_ContinuumLimit_DM} except that there are extra macroscopic exchange terms. These macroscopic exchange terms depend on the collisional invariants of the Vicsek operator.

\subsection{The equilibria of the Vicsek operators} 

We suppose that $\delta = O(1)$ and we want to take the limit $\eps \rightarrow 0$ in system \eqref{Eq:f0_eps_delta}-\eqref{Eq:f1_eps_delta}. Therefore, assuming that the distribution function $f_{0}^{\eps}$ and $f_{1}^{\eps}$ converge to limits denoted by $f_{0}$ and $f_{1}$ as $\eps \rightarrow 0$, these limits satisfy the equilibria : 
\begin{equation*}
Q_{0}(f_{0}) = 0\quad \text{ and }\quad Q_{1}(f_{1}) = 0.
\end{equation*}
According to lemma 4.2 in \cite{2008_ContinuumLimit_DM}, the kernels of the Vicsek operators $Q_{0,1}$ are two-dimensional manifolds: there exists $\rho_{0,1}(\bx,t) \in \R$ and $\bOmega_{0,1}(\bx,t) \in \S^1$ such that distributions $f_{0}$ and $f_{1}$ equals
\begin{equation}
f_{0}(\bx,\bomega,t) = \rho_{0}(\bx,t)M_{\lambda_0,\bOmega_{0}(\bx,t)}(\bomega),\quad f_{1}(\bx,\bomega,t) = \rho_{1}(\bx,t)M_{\lambda_1,\bOmega_{1}(\bx,t)}(\bomega),
\label{Eq:local_eq}
\end{equation}
where $M_{\lambda,\bOmega}$ denotes the Von Mises distribution defined in eq. \eqref{Eq:vonMises}. We actually note that $\rho_{0,1}(,\bx ,t)$ and  $\bOmega_{0,1}(\bx,t)$ are the density and the mean direction of the equilibria distribution function:
\begin{align}
&\rho[f_{0,1}](\bx ,t) =\rho_{0,1}(\bx ,t),\label{Eq:moment1_eq_local}\\
&\bj[f_{0,1}](\bx ,t) = c_{0,1}\rho_{0,1}(\bx ,t)\bOmega_{0,1}(\bx ,t),\quad \bOmega[f_{0,1}](\bx ,t) = \bOmega_{0,1}(\bx ,t), \label{Eq:moment2_eq_local}
\end{align}
where $c_{0,1}$ are constants defined as:
\begin{equation}
c_{0,1} = \langle\cos\theta\rangle_{M_{\lambda_{0,1}}}.
\label{Eq:const_c}
\end{equation} 
For any function $s(\cos\theta)$, the brackets $\langle s(\cos\theta)\rangle_{M_\lambda}$ will denote the average of $s(\bomega\cdot\bOmega)$ with respect to $M_{\lambda,\bOmega}$:
\begin{equation}
\langle s(\cos\theta)\rangle_{M_{\lambda}} = \int_{\theta = 0}^{2\pi} s(\cos\theta)C_{\lambda}e^{\frac{\cos\theta}{\lambda}}d\theta.
\label{Eq:notation_bracket}
\end{equation}
Note that this definition is independent of $\bOmega$.

\subsection{The generalised collisional invariants}
\label{sec:gen_coll_inv}

 We would like now to obtain the dynamics of the macroscopic quantities, $\rho_{0,1}(\bx ,t)$ and  $\bOmega_{0,1}(\bx,t)$. With this aim, the usual method is to integrate equations (\ref{Eq:f0_eps_delta}-\ref{Eq:f1_eps_delta}) against collisional invariants, which are velocity functions $I(\bomega)$ belonging to the orthogonal of the image of $Q_{0,1}$. A condition to recover the dynamics of the equilibria is that the dimension of the vector space of collisional invariants equals the dimension of the vector space of local equilibria, which here is $2$. It is not the case in the Vicsek dynamics since the only known collisional invariant is mass: $I_1(\bomega) = 1$. In \cite{2008_ContinuumLimit_DM}, this difficulty is overcome by considering generalised collisional invariants $I(\bomega)$, that are collisional invariants valid only for the subset of the distribution functions with a prescribed mean direction:
\begin{equation*}
\int_{\bomega \in \S^{1}} Q_{0,1}(f)(\bx, \bomega, t)I(\bomega)\ d\bomega = 0,\quad \forall f \text{ such that } \bOmega[f] = \bOmega. 
\end{equation*}
In dimension 2, it has been shown in \cite{2008_ContinuumLimit_DM} that the vector space of the generalised collisional invariants for $Q_0$ (resp. $Q_1$) is spanned by $I_1$ and a function $(I_{2})_0$ (resp. $(I_{2})_1$), which is the unique solution with zero average of the elliptic equation:
\begin{equation}
\label{Eq:elliptic}
\partial_{\theta}\left(e^{\cos\theta /\lambda_{0,1}}\partial_{\theta}(I_{2})_{0,1}\right) = \sin\theta e^{\cos\theta /\lambda_{0,1}},
\end{equation}
where we identified the functions on $\S^1$ and the $2\pi$-periodic functions of $\R$, using polar coordinates in the basis $(\bOmega,\bOmega^{\perp})$. Their explicit expressions are:
\begin{equation}
(I_{2})_{0,1}(\theta) = \lambda_{0,1} \left(\pi \frac{\int_{0}^{\theta} e^{- \cos\varphi /\lambda_{0,1}}d\varphi}{\int_{0}^{\pi} e^{- \cos\varphi /\lambda_{0,1}}d\varphi} - \theta\right).\label{Eq:I2}
\end{equation}
We would like here to stress that these second generalized invariants $(I_{2})_{0,1}$ depend on the parameter $\lambda_{0,1} = d_{0,1}/\nu_{0,1}$ and then are a priori different for the two Vicsek operators: this would not be the case if  the model conserved momentum. To simplify the following computations, let us introduce the function $h_{0,1}\left(\cos\theta\right) = (I_{2})_{0,1}(\theta)/\sin\theta$. We thus have: 
\begin{equation}
(I_{2})_{0,1}(\bomega) = h_{0,1}\left(\bomega\cdot\bOmega[f]\right) \left(\bOmega[f]^{\perp}\cdot\bomega\right).
\label{Eq:h}
\end{equation}
 
\subsection{The macroscopic system}
\label{subsec:macro_syst}

As announced in the previous section, to obtain the dynamics of the densities $\rho_{0,1}(\bx ,t)$ and the mean directions $\bOmega_{0,1}(\bx,t)$ introduced in eq. \eqref{Eq:local_eq}, we now integrate system \eqref{Eq:f0_eps_delta}-\eqref{Eq:f1_eps_delta} with respect to the velocity variable $\bomega$, after having multiplied it by the collisional invariants $I_1$ and $I_2$.

\paragraph{Mass equations.} Integrating equations \eqref{Eq:f0_eps_delta}-\eqref{Eq:f1_eps_delta} (multiplied by the collisional invariant $I_{1}(\bomega) = 1$) with respect to $\bomega$, we easily obtain:
\begin{align}
\partial_{t} \rho[f_{0}^{\eps}] &= \frac{1}{\delta}\int_{\bomega \in \S^1} E(f_{0}^{\eps},f_{1}^{\eps}) d\bomega,\label{Eq:rho0_eps}\\
\partial_{t} \rho[f_{1}^{\eps}] + \nabla_{\bx}\cdot j[f_{1}^{\eps}] &= - \frac{1}{\delta}\int_{\bomega \in \S^1} E(f_{0}^{\eps},f_{1}^{\eps}) d\bomega. \label{Eq:rho1_eps}
\end{align}
So, in the limit $\eps \rightarrow 0$, using the relations (\ref{Eq:moment1_eq_local}-\ref{Eq:moment2_eq_local}) , system (\ref{Eq:rho0_eps})-(\ref{Eq:rho1_eps}) results in the following mass equations:
\begin{align}
\partial_{t} \rho_{0} &= \frac{1}{\delta}R(\rho_0,\bOmega_0,\rho_1,\bOmega_1),\label{Eq:rho0}\\
\partial_{t} \rho_{1} + \nabla_{\bx}\cdot (c_{1}\rho_{1}\bOmega_{1}) &= - \frac{1}{\delta}R(\rho_0,\bOmega_0,\rho_1,\bOmega_1),\label{Eq:rho1}
\end{align}
where the macroscopic exchange operator $R$ is derived from the microscopic one $E$ (defined in \eqref{Eq:Exchange_op}-\eqref{Eq:g_eps2}) and easy computations provides the following expression:
\begin{align}
R(\rho_0,\bOmega_0,\rho_1,\bOmega_1) &= \int_{\bomega\in\S^{1}} E(\rho_0 M_{\lambda_0,\bOmega_0},\rho_1 M_{\lambda_1,\bOmega_1}) d\bomega\nonumber\\
&= \tau_{1}\rho_{1} - \tau_{0}\rho_{0}  + \alpha\left(\tau_{1} - \tau_{0}\right)\rho_{0}\rho_{1}\Phi,\label{Eq:exchange_macro}
\end{align}
and $\Phi$ is the macroscopic alignment of the two populations:
\begin{equation}
\Phi = \frac{1 + c_{0}c_{1}(\bOmega_{0}\cdot \bOmega_{1})}{2}.
\label{Eq:Local_Align}
\end{equation}
System (\ref{Eq:rho0}-\ref{Eq:rho1}-\ref{Eq:exchange_macro}) exactly corresponds to equations (\ref{Eq:rho0_thm_eps}-\ref{Eq:rho1_thm_eps}) of theorem \ref{thm:limit_eps}.

\paragraph{Momentum equations.} Let us now multiply system (\ref{Eq:f0_eps_delta}-\ref{Eq:f1_eps_delta}) by the second generalised collisional invariants $(I_2)_{0,1}(\bomega)$ and then integrate with respect to $\bomega$. After some computations reported in appendix~\ref{annex:momentum_computations} and similar to those made in \cite{2008_ContinuumLimit_DM}, we obtain in the limit $\eps \rightarrow 0$: 
\begin{align}
\lambda_{0}^{-1}\langle(\sin\theta)^{2}h_{0}\rangle_{M_{\bOmega_{0}}}\rho_{0}\partial_{t}\bOmega_{0} &= \nonumber\\
&\hspace{-2cm} (\text{\bf Id}-\bOmega_{0}\otimes\bOmega_{0})\left[-\langle(\sin\theta)^{2}h_{0}\rangle_{M_{\bOmega_{0}}}\nabla_{\bx}\rho_{0} + \frac{1}{\delta}\bS_{0}(\rho_0,\bOmega_0,\rho_1,\bOmega_1)\right],\label{Eq:rhoOmega0}\\
\lambda_{1}^{-1}\langle(\sin\theta)^{2}h_{1}\rangle_{M_{\bOmega_{1}}}\rho_{1}\partial_{t}\bOmega_{1} &+ \lambda_{1}^{-1}\langle(\sin\theta)^{2}\cos\theta h_{1}\rangle_{M_{\bOmega_{1}}}\rho_{1}(\bOmega_{1}\cdot\nabla_{x})\bOmega_{1} = \nonumber\\ 
&\hspace{-2cm} (\text{\bf Id}-\bOmega_{1}\otimes\bOmega_{1})\left[-\langle(\sin\theta)^{2}h_{1}\rangle_{M_{\bOmega_{1}}}\nabla_{\bx}\rho_{1} + \frac{1}{\delta}\bS_{1}(\rho_0,\bOmega_0,\rho_1,\bOmega_1)\right],\label{Eq:rhoOmega1}
\end{align}
where $\bS_{0}$ and $\bS_{1}$ are the terms coming from the exchange operator: 
\begin{align}
\bS_{0}(\rho_0,\bOmega_0,\rho_1,\bOmega_1) &= \int_{\bomega\in\S^{1}} \ E(\rho_0 M_{\lambda_0,\bOmega_0},\rho_1 M_{\lambda_1,\bOmega_1})h_{0}(\bomega\cdot\bOmega_{0})\bomega d\bomega, \label{Eq:S_0}\\
\bS_{1}(\rho_0,\bOmega_0,\rho_1,\bOmega_1) &= \int_{\bomega\in\S^{1}} -E(\rho_0 M_{\lambda_0,\bOmega_0},\rho_1 M_{\lambda_1,\bOmega_1})h_{1}(\bomega\cdot\bOmega_{1})\bomega d\bomega. \label{Eq:S_1}
\end{align}
where the exchange operator $E$, defined in \eqref{Eq:Exchange_op}-\eqref{Eq:g_eps2}, can be expressed as function of the macroscopic quantities:
\begin{align*}
E(\rho_0 M_{\lambda_0,\bOmega_0},\rho_1 M_{\lambda_1,\bOmega_1}) &= - \tau_{0}g_1(\rho_1 ,\bOmega_1)\rho_0 M_{\lambda_0,\bOmega_0} + \tau_{1}g_0(\rho_0 ,\bOmega_0)\rho_1 M_{\lambda_1,\bOmega_1},\\
\intertext{with:}
g_{0,1}(\rho_{0,1} ,\bOmega_{0,1})(\bx, \bomega ,t) &=  g[\rho_{0,1} M_{\lambda_{0,1},\bOmega_{0,1}}](\bx, \bomega ,t)\\ 
&= 1 + \alpha \rho_{0,1}(\bx, t)\left(\frac{1 + c_{0,1}(\bomega\cdot\bOmega_{0,1}(\bx, t))}{2}\right).
\end{align*}
Note that the index of the functions $g_{0,1}$ refers to the parameters $\lambda_{0,1}$ and then to the constant $c_{0,1}$ appearing when computing the flux $j[\rho_{0,1} M_{\lambda_{0,1},\bOmega_{0,1}}]$ of the local equilibrium (in \eqref{Eq:g_eps2}).

Unlike the exchange terms in the mass equations (\ref{Eq:rho0}-\ref{Eq:rho1}), the operators $\bS_{0}$ and $\bS_{1}$ are not opposite because of the presence of $h_{0}$ and $h_{1}$: according to their definitions \eqref{Eq:I2}-\eqref{Eq:h}, $h_{0}$ and $h_{1}$ are different as soon as $\lambda_0 \neq \lambda_1$ . However, the operators $\bS_{0}$ and $\bS_{1}$ are still symmetric in subscripts: when changing subscripts $0$ to $1$ and $1$ to $0$ in $\bS_{0}$, we obtain $\bS_{1}$. 

After introducing the following constants:
\begin{equation}
\gamma_{1} = \frac{\langle(\sin\theta)^{2}\cos\theta h_{1}\rangle_{M_{\bOmega_{1}}}}{\langle(\sin\theta)^{2}h_{1}\rangle_{M_{\bOmega_{1}}}},\ \beta_{0,1} =\left(\langle(\sin\theta)^{2}h_{0,1}\rangle_{M_{\bOmega_{0,1}}}\right)^{-1},
\label{Eq:const_gamma_beta}
\end{equation}
system \eqref{Eq:rhoOmega0}-\eqref{Eq:rhoOmega1} can be also written as follows:
\begin{align}
&\rho_{0}\partial_{t}\bOmega_{0} = \lambda_{0}(\text{\bf Id}-\bOmega_{0}\otimes\bOmega_{0})\left[-\nabla_{\bx}\rho_{0} + \frac{1}{\delta}\beta_{0}\bS_{0}(\rho_0,\bOmega_0,\rho_1,\bOmega_1)\right],\label{Eq:rhoOmega0_prop}\\
&\rho_{1}\partial_{t}\bOmega_{1} + \gamma_{1}\rho_{1}(\bOmega_{1}\cdot\nabla_{x})\bOmega_{1} = \lambda_{1}(\text{\bf Id}-\bOmega_{1}\otimes\bOmega_{1})\left[-\nabla_{\bx}\rho_{1} + \frac{1}{\delta}\beta_{1}\bS_{1}(\rho_0,\bOmega_0,\rho_1,\bOmega_1)\right],\label{Eq:rhoOmega1_prop}
\end{align}
which are the mean direction equations \eqref{Eq:rhoOmega0_thm_eps}-\eqref{Eq:rhoOmega1_thm_eps} given in theorem~\ref{thm:limit_eps}.

\section{Large scale speed change dynamics: limit $\delta \rightarrow 0$}
\label{Sec:LargeScale_speedchange}

In this section, we provide the proof of theorem~\ref{thm:limit_delta}: we first work out the equilibria of the exchange operators, $R$, $\bS_0$ and $\bS_{1}$, appearing in the right-hand-side of system (\ref{Eq:rho0_thm_eps}-\ref{Eq:rhoOmega1_thm_eps}). Note that they are non linear operators acting on the macroscopic densities and mean directions and note also that this is the projected momentum operators $(\text{\bf Id}-\bOmega_{0}\otimes\bOmega_{0})\bS_0$ and $(\text{\bf Id}-\bOmega_{1}\otimes\bOmega_{1})\bS_1$ that are acting in the momentum equations. However, we will prove that the only equilibria of the momentum exchange operators, $\bS_0$ and $\bS_{1}$, are given by $\bOmega_0 = \bOmega_1$ or $\bOmega_0 = - \bOmega_1$. We then obtained the macroscopic system (\ref{Eq:rho_thm_delta}-\ref{Eq:Omega_thm_delta}) by employing an Hilbert expansion with respect to $\delta$ and thanks to a ``discrete'' version of the generalised collisional invariants methodology, we close the system.

\subsection{Equilibria for the densities} 
\label{sec:eq_density}

Let us first explicitly re-introduce the dependency of the macroscopic variables with respect to $\delta$: the solutions of system (\ref{Eq:rho0_thm_eps}-\ref{Eq:rhoOmega1_thm_eps}) with $\delta > 0$ are denoted $\rho_{0,1}^\delta(\bx,t)$ and $\bOmega_{0,1}^\delta(\bx,t)$. Supposing that $\rho_{0,1}^\delta$ and $\bOmega_{0,1}^\delta$ converge to $\rho_{0,1}$  and $\bOmega_{0,1}$ as $\delta$ goes to $0$ and taking this limit in mass equations (\ref{Eq:rho0_thm_eps})-(\ref{Eq:rho1_thm_eps}), we formally obtain:
\begin{equation*}
R(\rho_0,\bOmega_0,\rho_1,\bOmega_1) = 0,
\end{equation*}
From the definition of $R$ (eq. \eqref{Eq:exchange_macro}), we easily obtain the balance equation for the densities
\begin{equation}
\rho_{0} = f_{\Phi}(\rho_1),\quad f_{\Phi}(\rho_1) = \rho_1\left(\frac{\tau_1}{\tau_0} + \alpha\left(\frac{\tau_{1}}{\tau_{0}} - 1\right)\Phi\rho_{1}\right)^{-1},\label{Eq:density_balance}
\end{equation}
which can be equivalently expressed in terms of the total density $\rho = \rho_{0} + \rho_{1}$: 
\begin{equation}
\rho = k_{\Phi}(\rho_1),\quad k_{\Phi}(\rho_1) = \rho_1 + f_{\Phi}(\rho_1).\label{Eq:k}
\end{equation}
Note that in the general case, these relations are not explicit since the macroscopic alignment parameter $\Phi$ (defined in \eqref{Eq:Local_Align}) depends on the directions $\bOmega_0$ and $\bOmega_1$, whose evolutions are non-linearly related to the densities through mass and momentum equations (\ref{Eq:rhoOmega0_thm_eps})-(\ref{Eq:rhoOmega1_thm_eps}). This provides the first part of theorem~\ref{thm:limit_delta}.

Let us now make some remarks on this relation. Simple computations shows that the function $f_\Phi$ (and then function $k_\Phi$) is increasing. The following proposition gives the domain where both $\rho_0$ and $\rho_1$ are non negative.
\begin{prop} (Conditions for positivity) If $\alpha > 0$, $\tau_0 \neq \tau_1$ and supposing that $\rho_0$ and $\rho_1$ are non-negative, then the following results hold : 
\begin{enumerate}
\item if $\tau_1/\tau_0 > 1$, then $\rho_0$ is bounded:\quad $\rho_{0} \leqslant \displaystyle\frac{1}{\alpha\Phi\left(\tau_{1}/\tau_0 -1\right)} \leqslant \frac{1}{\alpha\Phi_{\min}\left(\tau_{1}/\tau_0 -1\right)},$ 
\item if $\tau_1/\tau_0 < 1$, then $\rho_1$ is bounded:\quad $\rho_{1} \leqslant \displaystyle\frac{1}{\alpha\Phi\left(\tau_{0}/\tau_1 - 1\right)} \leqslant \frac{1}{\alpha\Phi_{\min}\left(\tau_{0}/\tau_1 - 1\right)},$
\end{enumerate}
where $\Phi_{\min} = (1 - c_0 c_1)/2$ is the minimal value of the macroscopic alignment parameter.
\end{prop}
\noindent The proof of this proposition is easy and omitted. This proposition shows that the dependency of the exchange rates on the local alignment $\Phi$ implies that the population with the higher interaction frequency is bounded in time.

\subsection{Equilibria for the mean directions} 
\label{sec:eq_direction}

Secondly, as $\delta$ goes to $0$ in the mean direction equations (\ref{Eq:rhoOmega0_thm_eps})-(\ref{Eq:rhoOmega1_thm_eps}), the limit $\rho_{0,1}$  and $\bOmega_{0,1}$ of $\rho_{0,1}^\delta$ and $\bOmega_{0,1}^\delta$  satisfies:
\begin{equation}
(\text{\bf Id} - \bOmega_{0}\otimes\bOmega_{0})\bS_{0}(\rho_{0},\bOmega_0,\rho_{1},\bOmega_1) = 0,\quad (\text{\bf Id} - \bOmega_{1}\otimes\bOmega_{1})\bS_{1}(\rho_{0},\bOmega_0,\rho_{1},\bOmega_1) = 0. \label{Eq:momentum_balance}
\end{equation}
The most simple solutions of \eqref{Eq:momentum_balance} are $\bOmega_{0} =  \bOmega_{1}$ or $\bOmega_{0} =  - \bOmega_{1}$, i.e. when the two populations are in the same or in the opposite direction. Indeed, for instance in the case $\bOmega_{0} =  \bOmega_{1}$, considering the polar coordinates $\theta$ in the basis $(\bOmega_0,\bOmega_0^{\perp})$, we can easily checked from (\ref{Eq:S_0}) that the momentum exchange operator writes as follows:
\begin{align*}
&(\text{\bf Id} - \bOmega_{0}\otimes\bOmega_{0})\bS_{0}(\rho_{0},\bOmega_0,\rho_{1},\bOmega_0) = \int_{\theta = 0}^{2\pi} f_{0}(\cos\theta)\sin\theta d\theta\, \bOmega_{0}^{\perp},\\
&f_{0}(\cos\theta) = [ -\tau_{0}(1 + \frac{\alpha}{2} \rho_{1}(1 + c_{1}\cos\theta))\rho_0\exp(\cos\theta/\lambda_0)\\
&\hspace{2cm} + \tau_{1}(1 + \frac{\alpha}{2}\rho_{0}(1 + c_{0}\cos\theta))\rho_1 \exp(\cos\theta/\lambda_1)] h_{0}(\cos\theta),
\end{align*} 
and thus vanishes by even argument. Let us now explicit the expression of $(\text{\bf Id} -\bOmega_{0}\otimes\bOmega_{0})\bS_{0}$ in the general case:
\begin{align}
&(\text{\bf Id} -\bOmega_{0}\otimes\bOmega_{0})\bS_{0}(\rho_{0},\bOmega_0,\rho_{1},\bOmega_0) =\nonumber\\ 
&\hspace{1cm}- (\text{\bf Id} -\bOmega_{0}\otimes\bOmega_{0})(1+\alpha\frac{\rho_{1}}{2})\rho_{0}\tau_0\int_{\bomega\in\S^{1}}M_{\lambda_0,\bOmega_{0}}h_{0}(\bomega\cdot\bOmega_{0})\bomega d\bomega \nonumber\\
&\hspace{1cm}-(\text{\bf Id} -\bOmega_{0}\otimes\bOmega_{0})\alpha \frac{c_{1}\rho_1}{2}\rho_{0}\tau_0\int_{\bomega\in\S^{1}}(\bomega\otimes\bomega)M_{\lambda_0,\bOmega_{0}}h_{0}(\bomega\cdot\bOmega_{0})d\bomega \bOmega_{1}\nonumber\\
&\hspace{1cm} + (\text{\bf Id} -\bOmega_{0}\otimes\bOmega_{0})(1+\alpha\frac{\rho_{0}}{2})\rho_{1}\tau_{1}\int_{\bomega\in\S^{1}}M_{\lambda_1,\bOmega_{1}}h_{0}(\bomega\cdot\bOmega_{0})\bomega d\bomega \nonumber\\
&\hspace{1cm}+(\text{\bf Id} -\bOmega_{0}\otimes\bOmega_{0})\alpha \frac{c_{0}\rho_{0}}{2}\rho_1\tau_{1}\int_{\bomega\in\S^{1}}(\bomega\otimes\bomega)M_{\lambda_1,\bOmega_{1}}h_{0}(\bomega\cdot\bOmega_{0})d\bomega \bOmega_{0}.\nonumber
\end{align}
By even arguments, the first term vanishes and we can easily check that the integral in the second term can be written as follows:
\begin{align*}
&\int_{\bomega\in\S^{1}}(\bomega\otimes\bomega)M_{\lambda_0,\bOmega_{0}}h_{0}(\bomega\cdot\bOmega_{0})d\bomega =\\ &\hspace{2cm}\langle(\cos\theta)^{2}h_{0}\rangle_{M_{\lambda_{0}}}\ \bOmega_{0}\otimes\bOmega_{0} + \langle(\sin\theta)^{2}h_{0}\rangle_{M_{\lambda_{0}}}\ (\text{\bf Id} -\bOmega_{0}\otimes\bOmega_{0}).
\end{align*}
where the bracket notation refers to \eqref{Eq:notation_bracket}. We deduce the following simplified expression for the projection of $\bS_{0}$:
\begin{align}
&(\text{\bf Id} -\bOmega_{0}\otimes\bOmega_{0})\bS_{0}(\rho_{0},\bOmega_0,\rho_{1},\bOmega_0) =\nonumber\\ 
&\hspace{1cm}-\alpha \frac{c_{1}\rho_1}{2}\rho_{0}\tau_0\langle(\sin\theta)^{2}h_{0}\rangle_{M_{\lambda_{0}}}(\text{\bf Id} -\bOmega_{0}\otimes\bOmega_{0})\bOmega_{1}\nonumber\\
&\hspace{1cm}+ (1+\alpha\frac{\rho_{0}}{2})\rho_{1}\tau_{1}\int_{\bomega\in\S^{1}}M_{\lambda_1,\bOmega_{1}}h_{0}(\bomega\cdot\bOmega_{0})(\bomega\cdot\bOmega_{0}^{\perp}) d\bomega\, \bOmega_{0}^{\perp}\label{Eq:X_0_explicit}\\
&\hspace{1cm}+\alpha \frac{c_{0}\rho_{0}}{2}\rho_1\tau_{1}(\text{\bf Id} -\bOmega_{0}\otimes\bOmega_{0})\int_{\bomega\in\S^{1}}(\bomega\otimes\bomega)M_{\lambda_1,\bOmega_{1}}h_{0}(\bomega\cdot\bOmega_{0})d\bomega\, \bOmega_{0},\nonumber
\end{align}
where all the terms collinear to $\Omega_0$ was removed. An equivalent expression for $(\text{\bf Id} -\bOmega_{1}\otimes\bOmega_{1})\bS_1$ can be obtained. Note that the involved integrals include the products of functions of the variable $(\bomega\cdot\bOmega_0)$ times functions of the variable $(\bomega\cdot\bOmega_1)$. Therefore, equilibria different from $\bOmega_0 = \pm\bOmega_1$ might exist. However, the following proposition states that they are the only two possible solutions and thus completes the first statement of theorem~\ref{thm:limit_delta}.
\begin{prop} 
\label{Prop:momentum_balance}
The only solutions to equations (\ref{Eq:momentum_balance}) are given by the set:
\begin{equation*}
\left\{(\rho_{0},\bOmega_0,\rho_{1},\bOmega_1) \in (\R^{+}\times\S^1)^2,\ \bOmega_{0} =  \bOmega_{1}\text{ or }\bOmega_{0} =  - \bOmega_{1}\right\}.
\end{equation*}
\end{prop}
\noindent The proof of this proposition can be found in appendix~\ref{Appendix:momentum_balance}. This proposition proves that the only two equilibria are those where the two sub-population are locally directed in the same ($\bOmega_{0} =  \bOmega_{1}$) or in the opposite ($\bOmega_{0} =  -\bOmega_{1}$) direction. However, the question of the stability of the equilibria remains open and will be addressed in a future work.

\subsection{The large scale dynamics}

The results of the two previous sub-sections suggest that at equilibria, the dynamics of the two sub-populations could be described by the density of the whole population $\rho = \rho_0 + \rho_1$ and the common direction $\bOmega = \bOmega_0 = \pm \bOmega_1$. To recover these dynamics, let us consider the following expansions with respect to $\delta$:
\begin{align}
\rho_{0}^{\delta} &= \rho_{0} + \delta\tilde{\rho}_{0} + O(\delta^2),\quad   \rho_{1}^{\delta} = \rho_{1} + \delta\tilde{\rho}_{1} + O(\delta^2),\label{Eq:expansion_0}\\
\bOmega_{0}^{\delta} &= \bOmega_{0} + \delta\tilde\bOmega_{0} + O(\delta^2),\quad  \bOmega_{1}^{\delta}  = \bOmega_{1} + \delta\tilde\bOmega_{1} + O(\delta^2).\label{Eq:expansion_1}
\end{align}
Let us note that $\bOmega_{0}^{\delta}$ and $\bOmega_{0}$ are of norm $1$ and therefore $\tilde{\bOmega}_{0}$ is orthogonal to $\bOmega_{0}$ (resp. $\tilde{\bOmega}_{1}$ is orthogonal to $\bOmega_{1}$). Thus, we have the following expansions of the von-Mises distributions:
\begin{align*}
& M_{\lambda_0,\bOmega_0^{\delta}}(\bomega) = M_{\lambda_0,\bOmega_0}(\bomega)(1 + \delta\lambda_0^{-1}(\bomega\cdot\tilde\bOmega_0) + O(\delta^{2})),\\
& M_{\lambda_1,\bOmega_1^{\delta}}(\bomega) = M_{\lambda_1,\bOmega_1}(\bomega)(1 + \delta\lambda_1^{-1}(\bomega\cdot\tilde\bOmega_1) + O(\delta^{2})).
\end{align*}
Let us also provide the expansions of the exchange operators $R$, $\bS_0$ and $\bS_1$ with respect to $\delta$:
\begin{align*}
&R(\rho_{0}^{\delta},\bOmega_0^{\delta},\rho_{1}^{\delta},\bOmega_1^{\delta}) = R(\rho_{0},\bOmega_0,\rho_{1},\bOmega_1) + \delta  (DR)_{(\rho_{0},\bOmega_0,\rho_{1},\bOmega_1)}(\tilde\rho_{0},\tilde\bOmega_0,\tilde\rho_{1},\tilde\bOmega_1) + O(\delta^{2}),\\
&\bS_{0,1}(\rho_{0}^{\delta},\bOmega_0^{\delta},\rho_{1}^{\delta},\bOmega_1^{\delta}) = \bS_{0,1}(\rho_{0},\bOmega_0,\rho_{1},\bOmega_1) + \delta  (D\bS_{0,1})_{(\rho_{0},\bOmega_0,\rho_{1},\bOmega_1)}(\tilde\rho_{0},\tilde\bOmega_0,\tilde\rho_{1},\tilde\bOmega_1) + O(\delta^{2}),
\end{align*}
where $DR$, $D\bS_0$, and  $D\bS_1$ denote the linearised operators, whose expressions are provided by the following lemma.
\begin{lem}  
\label{Prop:expansion_delta} 
The linearised exchange operators are given by:
\begin{align*}
(DR)_{(\rho_{0},\bOmega_0,\rho_{1},\bOmega_1)}(\tilde\rho_{0},\tilde\bOmega_0,\tilde\rho_{1},\tilde\bOmega_1) =\ & \tau_1\tilde\rho_1 - \tau_0\tilde\rho_0\\
&+ \alpha\left(\tau_1 - \tau_0\right)(\tilde\rho_0\rho_1 + \rho_0\tilde\rho_1)\left(\frac{1 + c_0c_1 \bOmega_0\cdot\bOmega_1}{2}\right)\\
&+ \alpha\left(\tau_1 - \tau_0\right)\rho_0\rho_1\frac{c_0c_1}{2}\left(\tilde\bOmega_0\cdot\bOmega_1 + \bOmega_0\cdot\tilde\bOmega_1\right).\\
(D\bS_0)_{(\rho_{0},\bOmega_0,\rho_{1},\bOmega_1)}(\tilde\rho_{0},\tilde\bOmega_0,\tilde\rho_{1},\tilde\bOmega_1) =\ & \tau_1 \bX_{0} - \tau_{0} \bY_{0},
\end{align*}
where $\bX_0$ and $\bY_{0}$ stand for:
\begin{align*}
\bX_{0} =&  \int_{\bomega\in\S^{1}} \left[(Dg_{0})_{(\rho_0 ,\bOmega_0)}(\tilde\rho_{0} ,\tilde\bOmega_{0})\right]\rho_1 M_{\lambda_1,\bOmega_1}h_{0}(\bomega\cdot\bOmega_{0})\bomega d\bomega \\
&+\int_{\bomega\in\S^{1}} g_0(\rho_0 ,\bOmega_0)\left[(D\rho M_{\lambda_{1},\bOmega})_{(\rho_{1} ,\bOmega_{1})}(\tilde\rho_{1} ,\tilde\bOmega_{1})\right]h_{0}(\bomega\cdot\bOmega_{0})\bomega d\bomega\\
&+ \int_{\bomega\in\S^{1}} g_0(\rho_0 ,\bOmega_0)\, \rho_1 M_{\lambda_1,\bOmega_1}\, h_{0}'(\bomega\cdot\bOmega_{0})(\bomega\otimes\bomega) d\bomega\ \tilde\bOmega_0 ,\\
\bY_{0} =&  \int_{\bomega\in\S^{1}} \left[(Dg_{1})_{(\rho_1 ,\bOmega_1)}(\tilde\rho_{1} ,\tilde\bOmega_{1})\right]\rho_0 M_{\lambda_0,\bOmega_0}h_{0}(\bomega\cdot\bOmega_{0})\bomega d\bomega\\
& + \int_{\bomega\in\S^{1}} g_1(\rho_1 ,\bOmega_1)\left[(D\rho M_{\lambda_{0},\bOmega})_{(\rho_{0} ,\bOmega_{0})}(\tilde\rho_{0} ,\tilde\bOmega_{0})\right]h_{0}(\bomega\cdot\bOmega_{0})\bomega d\bomega\\
&+ \int_{\bomega\in\S^{1}} g_1(\rho_1 ,\bOmega_1)\,  \rho_0 M_{\lambda_0,\bOmega_0}\, h_{0}'(\bomega\cdot\bOmega_{0})(\bomega\otimes\bomega) d\bomega\ \tilde\bOmega_0 ,
\end{align*}
and $(Dg_{0,1})$ and $(D\rho M_{\lambda_{0,1},\bOmega})$ denote the linearisation of the operators $g_{0,1}$ and $(\rho,\Omega) \mapsto \rho M_{\lambda_{0,1},\bOmega}$ and are given by:
\begin{align*}
&(Dg_{0,1})_{(\rho ,\bOmega)}(\tilde\rho ,\tilde\bOmega) = \frac{\alpha}{2} \left(\tilde\rho + c_{0,1}(\tilde\rho\bOmega + \rho\tilde\bOmega)\cdot\bomega\right),\\
&(D\rho M_{\lambda_{0,1},\bOmega})_{(\rho ,\bOmega)}(\tilde\rho ,\tilde\bOmega) = \left(\tilde\rho  + \rho\lambda_{0,1}^{-1}(\bomega\cdot\tilde\bOmega)\right)M_{\lambda_{0,1},\bOmega}.
\end{align*}
The expression of $(D\bS_1)$ is just obtained by changing the index $0$ into $1$ and $1$ into $0$.  
%
\end{lem}
\noindent The proof of this lemma is elementary and is omitted. Let us note that the expansion of the projected momentum exchange operator is as follows:
\begin{align}
(\text{\bf Id} - \bOmega_0^{\delta}\otimes\bOmega_0^{\delta})\bS_0(\rho_{0}^{\delta},\bOmega_0^{\delta},\rho_{1}^{\delta},\bOmega_1^{\delta}) =\ & (\text{\bf Id} - \bOmega_0\otimes\bOmega_0)\bS_0(\rho_{0},\bOmega_0,\rho_{1},\bOmega_1)\nonumber\\
&\hspace{-2cm} + \delta\, (\text{\bf Id}-\bOmega_0\otimes\bOmega_0)\left[(D\bS_{0,1})_{(\rho_{0},\bOmega_0,\rho_{1},\bOmega_1)}(\tilde\rho_{0},\tilde\bOmega_0,\tilde\rho_{1},\tilde\bOmega_1)\right]\nonumber\\
&\hspace{-2cm} -\delta\, (\tilde\bOmega_0\cdot \bS_0(\rho_{0},\bOmega_0,\rho_{1},\bOmega_1))\bOmega_0\nonumber\\
&\hspace{-2cm} - \delta(\bOmega_0\cdot\bS_0(\rho_{0},\bOmega_0,\rho_{1},\bOmega_1))\tilde\bOmega_0 + O(\delta^2).\label{Eq:projection_expansion}
\end{align}

\paragraph{Hilbert expansion.} We then insert expansions \eqref{Eq:expansion_0}-\eqref{Eq:expansion_1} into equations (\ref{Eq:rho0_thm_eps})-(\ref{Eq:rho1_thm_eps})-(\ref{Eq:rhoOmega0_thm_eps})-(\ref{Eq:rhoOmega1_thm_eps}).

$\bullet$ At the leading order $O(\delta^{-1})$, we obviously obtain that $(\rho_{0},\bOmega_0,\rho_{1},\bOmega_1)$ are equilibria of the exchange operators:
\begin{align}
&R(\rho_{0},\bOmega_0,\rho_{1},\bOmega_1) = 0, \\
&(\text{\bf Id} - \bOmega_{0}\otimes\bOmega_{0})\bS_0(\rho_{0},\bOmega_0,\rho_{1},\bOmega_1) = 0, \label{Eq:Omega0_equi}\\
&(\text{\bf Id} - \bOmega_{1}\otimes\bOmega_{1})\bS_1(\rho_{0},\bOmega_0,\rho_{1},\bOmega_1) = 0. \label{Eq:Omega1_equi}
\end{align}
Consequently, according to previous sections \ref{sec:eq_density} and \ref{sec:eq_direction}, the following relations hold:
\begin{align*}
&\rho = k_{\Phi}(\rho_1),\quad \bOmega_0 = \bOmega_1,\quad\text{ with  }\Phi = (1 + c_0c_1)/2,\\
\text{ or }\quad &\rho = k_{\Phi}(\rho_1),\quad \bOmega_0 = - \bOmega_1,\quad\text{ with  }\Phi = (1 - c_0c_1)/2.
\end{align*}
where $\rho = \rho_0 + \rho_1$ is the total density and the function $k_{\Phi}$ is defined by (\ref{Eq:k}). Therefore, we would like to stress that the very simple form of these equilibria is essentially due to the fact that both mean directions equilibria and density equilibria are characterised by the alignment of the mean directions.

$\bullet$ At order $O(1)$, system (\ref{Eq:rho0_thm_eps})-(\ref{Eq:rho1_thm_eps})-(\ref{Eq:rhoOmega0_thm_eps})-(\ref{Eq:rhoOmega1_thm_eps}) becomes:
\begin{align}
\hspace{-1.5cm}\partial_{t}\rho_{0} &= (DR)_{(\rho_{0},\bOmega_0,\rho_{1},\bOmega_1)}(\tilde\rho_{0},\tilde\bOmega_0,\tilde\rho_{1},\tilde\bOmega_1),\label{Eq:rho0_order1}\\[4pt]
\hspace{-1.5cm}\partial_{t} \rho_{1} + \nabla_{\bx}\cdot (c_{1}\rho_{1}\bOmega_{1}) &= - (DR)_{(\rho_{0},\bOmega_0,\rho_{1},\bOmega_1)}(\tilde\rho_{0},\tilde\bOmega_0,\tilde\rho_{1},\tilde\bOmega_1),\label{Eq:rho1_order1}
\end{align}
\begin{align}
\rho_{0}\partial_{t}\bOmega_{0} +\ & \lambda_{0}(\text{\bf Id}-\bOmega_{0}\otimes\bOmega_{0})\nabla_{\bx}\rho_{0} = \nonumber\\
&\lambda_{0}\beta_{0}(\text{\bf Id}-\bOmega_{0}\otimes\bOmega_{0})\left[(D\bS_0)_{(\rho_{0},\bOmega_0,\rho_{1},\bOmega_1)}(\tilde\rho_{0},\tilde\bOmega_0,\tilde\rho_{1},\tilde\bOmega_1)\right]\nonumber\\
& - \lambda_{0}\beta_0(\bOmega_0\cdot \bS_0(\rho_{0},\bOmega_0,\rho_{1},\bOmega_1))\tilde\bOmega_0,\label{Eq:Omega0_order1}\\[8pt]
\rho_{1}\partial_{t}\bOmega_{1} +\ &  \gamma_{1}\rho_{1}(\bOmega_{1}\cdot\nabla_{\bx})\bOmega_{1} + \lambda_{1}(\text{\bf Id}-\bOmega_{1}\otimes\bOmega_{1})\nabla_{\bx}\rho_{1} =\nonumber\\
&\lambda_{1}\beta_{1}(\text{\bf Id}-\bOmega_{1}\otimes\bOmega_{1})\left[(D\bS_1)_{(\rho_{0},\bOmega_0,\rho_{1},\bOmega_1)}(\tilde\rho_{0},\tilde\bOmega_0,\tilde\rho_{1},\tilde\bOmega_1)\right]\nonumber\\
& - \lambda_{1}\beta_1(\bOmega_1\cdot \bS_1(\rho_{0},\bOmega_0,\rho_{1},\bOmega_1))\tilde\bOmega_1,\label{Eq:Omega1_order1}
\end{align}
where $DR$, $D\bS_0$, $D\bS_1$ are the linearised exchange operators, whose expressions are given in lemma~\ref{Prop:expansion_delta}. Note that the terms $(\tilde\bOmega_0\cdot \bS_0)\bOmega_0$ (resp. $(\tilde\bOmega_1\cdot \bS_1)\bOmega_1$) in (\ref{Eq:projection_expansion}) do not appear in equation  (\ref{Eq:Omega0_order1}) (resp. (\ref{Eq:Omega1_order1})): indeed, according to \eqref{Eq:Omega0_equi} (resp. (\ref{Eq:Omega1_equi})), $\bS_0$ is parallel to $\bOmega_0$ (resp. $\bS_1$ is parallel to $\bOmega_1$). 

\paragraph{Closure.} System (\ref{Eq:rho0_order1})-(\ref{Eq:rho1_order1})-(\ref{Eq:Omega0_order1})-(\ref{Eq:Omega1_order1}) is not closed: it depends on the dynamics of the first order correction terms ($\tilde \rho_0$, $\tilde \rho_1$, $\tilde \bOmega_0$, $\tilde \bOmega_1$) that would be provided by the equations at order $O(\delta)$. However, by expressing compatibility conditions, we are able to close the system. We consider the case $\bOmega_0 = \bOmega_1$.

Actually,  adding the two density equations, we get the following closed equation:
\begin{equation*}
\partial_t (\rho_0 + \rho_1) + \nabla_{\bx}\cdot(c_1\rho_1\bOmega_1) = 0,
\end{equation*} 
where the exchange terms have been cancelled. 

For equations (\ref{Eq:Omega0_order1})-(\ref{Eq:Omega1_order1}), supposing that $\tilde\rho_{0}$ and $\tilde\rho_{1}$ are zero, the operator at the right-hand side of equations (\ref{Eq:Omega0_order1})-(\ref{Eq:Omega1_order1}) reduces to a linear operator in the two-dimensional space $\text{vect}(\bOmega_0^{\perp})\times\text{vect}(\bOmega_0^{\perp})$ acting on $\tilde\bOmega_0, \tilde\bOmega_1 \in \text{vect}(\bOmega_0^{\perp})$, where $\text{vect}(\bOmega_0^{\perp})$ denotes the line spanned by $\bOmega_0^{\perp}$. Its Null-Space is not reduced to $\left\{0\right\}$ since it is not the case for the original operator. Thus its image is one-dimensional (the operator is neither bijective nor zero) and a closed equation can be obtained just by expressing that the left-hand side of equations (\ref{Eq:Omega0_order1})-(\ref{Eq:Omega1_order1}) have to belong to the one-dimensional image of the linearised operator. Actually, an explicit expression of this linear operator can be obtained. 

\begin{prop}  \label{Prop:collisional_invariant}

The following identities hold: 
\begin{align*} 
&\forall \rho_0,\rho_1,\tilde\rho_0,\tilde\rho_1 \in \R^{+},\forall\, \bOmega_0 = \pm \bOmega_1 \in \mathbb{S}^{1},\ \forall\, \tilde\bOmega_0, \tilde\bOmega_1 \in \text{vect}(\bOmega_0^{\perp}) = \text{vect}(\bOmega_1^{\perp}),\\[4pt]
&\lambda_{0}\beta_{0}(\text{\bf Id}-\bOmega_{0}\otimes\bOmega_{0})\left[(D\bS_0)_{(\rho_{0},\bOmega_0,\rho_{1},\bOmega_1)}(\tilde\rho_{0},\tilde\bOmega_0,\tilde\rho_{1},\tilde\bOmega_1)\right]\\
&\qquad - \lambda_{0}\beta_0(\bOmega_0\cdot \bS_0(\rho_{0},\bOmega_0,\rho_{1},\bOmega_1))\tilde\bOmega_0 =  
\left\{ \begin{array}{ll} A_{0}^+(\rho_0,\rho_1)(\tilde\bOmega_1 - \tilde\bOmega_0)&\text{ if }\bOmega_0 = \bOmega_1,\\ A_{0}^-(\rho_0,\rho_1)(\tilde\bOmega_1 + \tilde\bOmega_0)&\text{ if }\bOmega_0 = -\bOmega_1, \end{array}\right. \\[4pt]
&\lambda_{1}\beta_{1}(\text{\bf Id}-\bOmega_{1}\otimes\bOmega_{1}) \left[(D\bS_1)_{(\rho_{0},\bOmega_0,\rho_{1},\bOmega_1)}(\tilde\rho_{0},\tilde\bOmega_0,\tilde\rho_{1},\tilde\bOmega_1)\right]\\ 
&\qquad - \lambda_{1}\beta_1(\bOmega_1\cdot \bS_1(\rho_{0},\bOmega_0,\rho_{1},\bOmega_1))\tilde\bOmega_1 =    
\left\{ \begin{array}{ll} A_{1}^+(\rho_0,\rho_1)(\tilde\bOmega_0 - \tilde\bOmega_1)&\text{ if }\bOmega_0 = \bOmega_1,\\ A_{1}^-(\rho_0,\rho_1)(\tilde\bOmega_1 + \tilde\bOmega_0)&\text{ if }\bOmega_0 = -\bOmega_1, \end{array}\right.
\end{align*}
where $A_0^+$ and $A_0^-$ are two bilinear functions with respect to $\rho_0$ and $\rho_1$ and whose expressions are given by:
\begin{align*}
&A_{0}^+(\rho_0,\rho_1) = \lambda_{0}\beta_{0}\left(\tau_1(1+\alpha\frac{\rho_0}{2})\rho_1\lambda_1^{-1}\langle\sin^2\theta h_0\rangle_{M_{\lambda_1}} + \tau_1\alpha \frac{c_0\rho_0}{2}\rho_1\lambda_1^{-1}\langle\sin^2\theta\cos\theta h_0\rangle_{M_{\lambda_1}}\right.\\
&\hspace{2.5cm}\left. - \tau_0\alpha \frac{c_1\rho_1}{2}\rho_0\langle\sin^2\theta h_0\rangle_{M_{\lambda_0}}\right),\\
&A_{0}^-(\rho_0,\rho_1) = \lambda_{0}\beta_{0}\left(\tau_1(1+\alpha\frac{\rho_0}{2})\rho_1\lambda_1^{-1}\langle\sin^2\theta h_0^-\rangle_{M_{\lambda_1}} - \tau_1\alpha \frac{c_0\rho_0}{2}\rho_1\lambda_1^{-1}\langle\sin^2\theta\cos\theta h_0^-\rangle_{M_{\lambda_1}}\right.\\
&\hspace{2.5cm}\left. - \tau_0\alpha \frac{c_1\rho_1}{2}\rho_0\langle\sin^2\theta h_0\rangle_{M_{\lambda_0}}\right),
\end{align*}
and $A_1^+$ and $A_1^-$ with the same expression by changing $1$ to $0$ and $0$ to $1$, and where, for any function $h$, $h^{-} : x \mapsto h(-x)$ denotes its symmetric. 
\end{prop}
\noindent The proof of this proposition is developed in appendix~\ref{Appendix:CollisionalInvariant}: it relies on the simplification coming from the equilibria relations $\bOmega_0 = \pm \bOmega_1$. Multiplying equation \eqref{Eq:Omega0_order1} by $A_0^+$ (resp. $A_0^-$) and equation \eqref{Eq:Omega1_order1} by $A_0^+$ (resp. $A_0^-$) and then adding them (resp. subtracting them),  it is an easy matter to derive the following closed system when equilibrium $\bOmega_0 = \bOmega_1$ (resp. $\bOmega_0 = -\bOmega_1$) is reached:
\begin{prop}\label{prop:macro_model} In the case where $\bOmega_0 = \bOmega_1$ (resp. $\bOmega_0 = -\bOmega_1$), system (\ref{Eq:rho0_order1})-(\ref{Eq:rho1_order1})-(\ref{Eq:Omega0_order1})-(\ref{Eq:Omega1_order1}) yields  the following closed system for the total density $\rho = \rho_0 + \rho_1$ and the common mean direction $\bOmega = \bOmega_0 = \bOmega_1$ (resp. $\bOmega = \bOmega_1 = -\bOmega_0$):
\begin{align}
&\partial_{t} \rho + \nabla_{x}\cdot (c_{1}\rho_1[\rho]\bOmega) = 0,\label{Eq:rho_equilibre}\\
&M^+(\rho) \partial_{t}\bOmega + \gamma_1 N^+(\rho)\ (\bOmega\cdot\nabla_{\bx})\bOmega = -P^+(\rho)(\text{\bf Id}-\bOmega\otimes\bOmega)\ \nabla_{\bx}\rho, \label{Eq:Omega_equilibre}\\
\left[resp.\ \right. &\left. M^-(\rho) \partial_{t}\bOmega + \gamma_1 N^-(\rho)\ (\bOmega\cdot\nabla_{\bx})\bOmega = -P^-(\rho)(\text{\bf Id}-\bOmega\otimes\bOmega)\ \nabla_{\bx}\rho,\right]
\end{align}
where functions $M^\pm, N^\pm, P^\pm$ are given by:
\begin{align*}
&M^\pm(\rho) = A_1^\pm[\rho]\rho_0[\rho] + A_0^\pm[\rho]\rho_1[\rho],\\
&N^\pm(\rho) = \rho_1[\rho] A_0^\pm[\rho],\\
&P^\pm(\rho) = \lambda_{0}A_1^\pm[\rho]\rho_0'[\rho] + \lambda_1 A_0^\pm[\rho]\rho_1'[\rho],
\end{align*}
where $A_0^\pm$, $A_1^\pm$, $\rho_0$ and $\rho_1$ are here considered as functions of $\rho$:
\begin{equation*}
\rho_1[\rho] = k_{\Phi}^{-1}(\rho),\ \rho_0[\rho] = \rho - k_{\Phi}^{-1}(\rho),\ A_{0,1}^\pm[\rho] =A_{0,1}^\pm(\rho_0[\rho],\rho_1[\rho])
\end{equation*}
with $k_{\Phi}$ given by (\ref{Eq:k}) and $A_0^\pm$ and $A_1^\pm$ given in proposition~\ref{Prop:collisionnal invariant}.
\end{prop}
\noindent System (\ref{Eq:rho_equilibre})-(\ref{Eq:Omega_equilibre}) provides the dynamics of the two populations at equilibrium. The study of its mathematical properties will be the subject of future work.
\\

\textbf{Remark:} The derivation of this closed system can be interpreted with a ``collisional invariant" viewpoint. To obtain the density equation, we use that the exchange interactions preserve mass and thus the vector $(R,-R)$ as well as its linearised operator $(\tilde R,-\tilde R)$ satisfies:
\begin{equation*}
\forall \rho_0,\rho_1 \in \R^{+},\ \forall \bOmega_0,\bOmega_1 \in \S^{1},\quad \left(\begin{array}{c} 1 \\ 1  \end{array}\right)\cdot\left(\begin{array}{c} R\\ - R \end{array}\right) = 0.
\end{equation*}
Since momentum is not conserved, such a relation is lacking for the momentum exchange operator $(\text{\bf Id}-\bOmega_0\otimes\bOmega_0)\bS_0,(\text{\bf Id}-\bOmega_1\otimes \bOmega_1)\bS_1$. However, the previous results show that the linearised operator around an equilibria have such a collisional invariant (equal to $(A_0,A_1)$) and that it is sufficient to conclude.

\section{Conclusion}

In this article, we have studied the dynamics at large time and space scale of a two-phase Vicsek model, in which particles speed can take only the two values $0$ or $1$. To this aim, we proceed in two steps. First, using generalised collision invariants \cite{2008_ContinuumLimit_DM}, we provides a two-phase Vicsek continuum model, where the densities and the directions dynamics are coupled through a speed change operator. Due to the non-conservativity of the model, the mean direction is not conserved. However, once figuring out the equilibria of the speed change operator, the averaged two-phase model can be obtained by integrating the dynamics against a vector orthogonal to the image of the linearised speed change operator. This leads to a new non-conservative model where coefficients non-linearly depend on the local mean density.

Future works would consist into investigating the mathematical properties of this new model (stability of equilibria, hyperbolicity). As for the Vicsek model \cite{2010_MotschLN_NumMacroVic}, appropriate numerical schemes would be also devised. Another point concerns the macroscopic dynamics of the two-phase model when speed changes are more frequent (or as frequent as) than Vicsek interactions: the question whether flocking occurs in that asymptotic regimes is still unknown.

The model considered in this article is designed to analyze alignment interactions during grazing period of mammal herds. However, the model is minimalist and several improvements should be investigated. First, long range attraction and short range repulsion interactions could be added to ensure cohesion of the group and to model congestion \cite{2010_CarilloForn_Review, 2010_DegondNavoret_Congestion}. Secondly, as in any living system, the assumption of homogeneous interactions among the animals is not satisfactory:  animals present different behaviours, which can be modelled by internal variables such as the degree of attention or the degree of hunger of each individual. Such heterogeneous behaviour has been already considered for cells population \cite{2008_BellomoDelitala_cells,2005_BellouDellitala}. Finally, we could also make the model depending on environment variables such as the local resource (the level of grass) or/and the topography. All these modeling refinements may affect the global herd dynamics.

\paragraph{Acknowledgements}  The author would like to express his gratitude to Pierre Degond and Francesco Ginelli for their fruitful suggestions. He wishes also to thank Richard Bon, Hugues Chat\'e, Marie-H\'el\`ene Pillot and Guy Th\'eraulaz for stimulating discussions. This work has been supported by the ``Agence Nationale de la Recherche'', in the frame of the contract PANURGE 07-BLAN-0208.

\appendix
\section{Appendix: Derivation of the momentum equations}
\label{annex:momentum_computations}

For the sake of completeness, we present here the derivation of the evolution equation of $\bOmega_1$, given by (\ref{Eq:rhoOmega1}). The derivation of equation (\ref{Eq:rhoOmega0}) is similar. 

We multiply system (\ref{Eq:f1_eps_delta}) by the generalised collisional invariant:
\begin{equation*}
(I_2)_{1}^{\eps} =h_1\left(\bomega\cdot\bOmega[f_1^{\eps}]\right) \left(\bOmega[f_1^{\eps}]^{\perp}\cdot\bomega\right),
\end{equation*}
introduced in section~\ref{sec:gen_coll_inv} (eq. \eqref{Eq:h}) and we integrate it with respect to $\bomega$. In the limit $\eps \rightarrow 0$, we have $\bOmega[f_1^{\eps}] \rightarrow \bOmega_1$ and $(I_2)_{1}^{\eps} \rightarrow h_1\left(\bomega\cdot\bOmega_1\right) \left(\bOmega_1^{\perp}\cdot\bomega\right)$. Therefore, we obtain:
\begin{align*}
&(\bOmega_1^{\perp}\otimes\bOmega_1^{\perp})\bX  = 0,\\
&\bX = \int_{\bomega \in \S^1} \left[\partial_{t}(\rho_1 M_{\lambda_1,\bOmega_1}) + \bomega\cdot\nabla_{\bx}(\rho_1 M_{\lambda_1,\bOmega_1}) + \frac{1}{\delta}E(\rho_0 M_{\lambda_0,\bOmega_0},\rho_1 M_{\lambda_1,\bOmega_1})\right] h_1 \bomega d\bomega, 
\end{align*}
where $h_1$ and $M_{\lambda_1\bOmega_1}$ are functions of $(\bomega\cdot\bOmega_1)$ and  $M_{\lambda_0,\bOmega_0}$ depends only on $(\bomega\cdot\bOmega_0)$. The derivative of $M_{\lambda_1,\bOmega_1}$ with respect to $\bOmega_1$ acting on a tangent vector $d\bOmega$ to the circle is given by:
\begin{equation*}
\frac{\partial M_{\lambda_1,\bOmega_1}}{\partial\bOmega_1}(d\bOmega) = \lambda_1^{-1}(\bomega\cdot\bOmega_1)(\bomega\cdot d\bOmega)M_{\lambda_1,\bOmega_1}.
\end{equation*}
Thus, we have:
\begin{align*}
\bX =\ & \int_{\bomega \in \S^1} \left[\partial_{t}\rho_1 + \bomega\cdot\nabla_{\bx}\rho_1 + \lambda_1^{-1}\rho_1\left(\bomega\cdot\partial_{t}\bOmega_1 + (\bomega\otimes\bomega):\nabla_{\bx}\bOmega_1\right)\right] M_{\bOmega_1}h_1 \bomega d\bomega\\
& - \frac{1}{\delta}\bS_1,
\end{align*}
where $S_1$ is given by (\ref{Eq:S_1}). The symbol ':' denotes the contracted product of two tensors (if $A = (A_{i,j})_{i,j=1,2}$ and $B = (B_{i,j})_{i,j=1,2}$ are two tensors then $A:B = \sum_{i,j = 1,2} A_{i,j}B_{i,j}$) and $\nabla_x\bOmega$ is the gradient tensor of the vector $\bOmega$: $(\nabla_x\bOmega)_{i,j} = \partial_{x_i}\bOmega_j$. The four first terms in this formula, denoted $\bX_1$ to $\bX_4$, are computed using the polar coordinate $\theta$ related to the Cartesian basis $(\bOmega_1,\bOmega_1^{\perp})$: in this basis $h_1$ and $M_{\lambda_1,\bOmega_1}$ are even functions depending on $\cos\theta$. 

By writing $\bomega = (\bomega\cdot\bOmega_1)\bOmega_1 + (\bomega\cdot\bOmega_1^{\perp})\bOmega_1^{\perp}$, we decompose $\bX_1$ as follows:
\begin{eqnarray*}
\bX_{1} &=& \int_{\bomega \in \S^1} \partial_{t}\rho_1 M_{\lambda_1,\bOmega_1}h_1\bomega d\bomega\\ 
&=& \partial_{t}\rho_1 \int_{\theta = -\pi}^{\pi} h_1 M_{\lambda_1,\bOmega_1}\cos\theta d\theta \ \bOmega_1 + \partial_{t}\rho_1  \int_{\theta = -\pi}^{\pi} h_1 M_{\lambda_1,\bOmega_1}\sin\theta d\theta \ \bOmega_1^{\perp},
\end{eqnarray*}
The second term is zero (the integration of an odd function on a symmetric interval) and thus we obtain:
\begin{equation}
(\bOmega_1^{\perp}\otimes\bOmega_1^{\perp}) \bX_{1} = 0.\label{Eq:X1}
\end{equation}
A similar computation shows that:
\begin{eqnarray*}
\bX_{2} &=& \int_{\bomega \in \S^1} \left[(\bomega\otimes\bomega)\nabla_{\bx}\rho_1\right] M_{\bOmega_1}h_1d\bomega\\
&=& \int_{\theta = -\pi}^{\pi} h_1 M_{\lambda_1,\bOmega_1}\cos^2\theta d\theta \ (\bOmega_1\otimes\bOmega_1)\nabla_{\bx}\rho_1\\
&& +\int_{\theta = -\pi}^{\pi} h_1 M_{\lambda_1,\bOmega_1}\cos\theta \sin\theta d\theta \ [(\bOmega_1\otimes\bOmega_1^{\perp}) + (\bOmega_1^{\perp}\otimes\bOmega_1)]\nabla_{\bx}\rho_1\\
&& + \int_{\theta = -\pi}^{\pi} h_1 M_{\lambda_1,\bOmega_1}\sin^2\theta d\theta \ (\bOmega_1^{\perp}\otimes\bOmega_1^{\perp})\nabla_{\bx}\rho_1.
\end{eqnarray*}
The first term is parallel to $\bOmega_1$ and the second term is zero. Therefore, we obtain:
\begin{equation}
(\bOmega_1^{\perp}\otimes\bOmega_1^{\perp}) \bX_{2} = \langle \sin^2\theta h_1\rangle_{M_{\lambda_1}}(\bOmega_1^{\perp}\otimes\bOmega_1^{\perp})\nabla_{\bx}\rho_1. \label{Eq:X2}
\end{equation}
With similar computations, we obtain the following results for $\bX_3$:
\begin{eqnarray}
\bX_{3} &=& \lambda_1^{-1}\rho_1\int_{\bomega \in \S^1} \left[(\bomega\otimes\bomega)\cdot\partial_{t}\bOmega_1\right] M_{\lambda_1,\bOmega_1}h_1 d\bomega,\nonumber\\
(\bOmega_1^{\perp}\otimes\bOmega_1^{\perp}) \bX_{3} &=& \lambda_1^{-1}\rho_1\langle \sin^2\theta h_1\rangle_{M_{\lambda_1}}(\bOmega_1^{\perp}\otimes\bOmega_1^{\perp})\partial_{t}\bOmega_1. \label{Eq:X3}
\end{eqnarray}
For $\bX_4$, we have:
\begin{align*}
\bX_{4} =\ & \lambda_1^{-1}\rho_1\int_{\bomega \in \S^1} \left[(\bomega\otimes\bomega):\nabla_{\bx}\bOmega_1\right]M_{\lambda_1,\bOmega_1} h_1\bomega d\bomega \\
=\ & \lambda_1^{-1}\rho_1\int_{\theta = -\pi}^{\pi} \left[\cos^2\theta(\bOmega_1\otimes\bOmega_1)+ \sin\theta\cos\theta(\bOmega_1\otimes\bOmega_1^{\perp}) \right.\\
&\hspace{1.5cm}\left. + \sin\theta\cos\theta(\bOmega_1^{\perp}\otimes\bOmega_1) + \sin^2\theta(\bOmega_1^{\perp}\otimes\bOmega_1^{\perp})\right]:\nabla_{\bx}\bOmega_1\ M_{\lambda_1,\bOmega_1}h_1 \cos\theta d\theta\ \bOmega_1\\
& + \lambda_1^{-1}\rho_1\int_{\theta = -\pi}^{\pi} \left[\cos^2\theta(\bOmega_1\otimes\bOmega_1) + \sin\theta\cos\theta(\bOmega_1\otimes\bOmega_1^{\perp}) \right.\\
&\hspace{1.5cm}\left. + \sin\theta\cos\theta(\bOmega_1^{\perp}\otimes\bOmega_1) + \sin^2\theta(\bOmega_1^{\perp}\otimes\bOmega_1^{\perp})\right]:\nabla_{\bx}\bOmega_1\ M_{\lambda_1,\bOmega_1}h_1 \sin\theta d\theta\ \bOmega^{\perp}.
\end{align*}
The odd function of $\theta$ vanishes and thus we are left with:
\begin{eqnarray*}
\bX_4 &=&\lambda_1^{-1}\rho_1\int_{\theta = -\pi}^{\pi} \left[\cos^2\theta(\bOmega_1\otimes\bOmega_1)+ \sin^2\theta(\bOmega_1^{\perp}\otimes\bOmega_1^{\perp})\right]:\nabla_{\bx}\bOmega_1\ M_{\lambda_1,\bOmega_1}h_1\cos\theta d\theta\ \bOmega_1\\
&& + \lambda_1^{-1}\rho_1\int_{\theta = -\pi}^{\pi} \sin\theta\cos\theta\left[ (\bOmega_1\otimes\bOmega_1^{\perp})  + (\bOmega_1^{\perp}\otimes\bOmega_1) \right]:\nabla_{\bx}\bOmega\ M_{\lambda_1,\bOmega_1}h_1\sin\theta d\theta\ \bOmega_1^{\perp}.
\end{eqnarray*}
Besides, we have for all $\bOmega \in \S^1$:
\begin{equation*}
(\bOmega\otimes\bOmega^{\perp}):\nabla_{\bx}\bOmega = \bOmega_{i}\bOmega^{\perp}_{j}\partial_{x_i}\bOmega_{j} = ((\bOmega\cdot\nabla_{\bx})\bOmega)\cdot\bOmega^{\perp},
\end{equation*}
and 
\begin{equation*}
(\bOmega^{\perp}\otimes\bOmega):\nabla_{\bx}\bOmega = \bOmega^{\perp}_{i}\bOmega_{j}\partial_{x_i}\bOmega_{j} = \frac{1}{2}\bOmega^{\perp}\cdot\nabla|\bOmega|^2 = 0.
\end{equation*}
So, we finally have
\begin{equation}
(\bOmega_1^{\perp}\otimes\bOmega_1^{\perp}) \bX_{4} = \lambda_1^{-1}\rho\langle \sin^2\theta\cos\theta h_1\rangle_{M_{\lambda_1}}(\bOmega_1^{\perp}\otimes\bOmega_1^{\perp})((\bOmega_1\cdot\nabla_{\bx})\bOmega_1).\label{Eq:X4}
\end{equation}
Equations (\ref{Eq:X1})-(\ref{Eq:X2})-(\ref{Eq:X3})-(\ref{Eq:X4}) results in the momentum equations (\ref{Eq:rhoOmega1}). 

\section[Appendix : The equilibria of the momenta operators]{Proof of proposition~\ref{Prop:momentum_balance} (Momenta balance)}
\label{Appendix:momentum_balance}

\begin{proof} We show here that if the equilibria condition holds:  
\begin{equation*}
(\text{\bf Id} - \bOmega_{0}\otimes\bOmega_{0})\bS_{0}(\rho_{0},\bOmega_0,\rho_{1},\bOmega_0) = 0
\end{equation*} then we have $\bOmega_0 = \pm \bOmega_1$. Let $\phi$ the angle between $\bOmega_{0}$ and $\bOmega_{1}$. We have the following computations.
\begin{enumerate} 
\item We can integrate expressions between $0$ and $\pi$ instead of $-\pi$ and $\pi$: 
\begin{align*}
&\int_{\bomega\in\S^{1}}M_{\lambda_1,\bOmega_{1}}h_{0}(\bomega\cdot\bOmega_{0})(\bomega\cdot\bOmega_{0}^{\perp}) d\bomega =\\ 
&\qquad\quad = C_{1}\int_{\theta = 0}^{\pi}\sin\theta (e^{\frac{1}{\lambda_{1}}\cos(\theta - \phi)} - e^{\frac{1}{\lambda_{1}}\cos(\theta + \phi)})h_{0}(\cos\theta)d\theta \\
&\qquad\quad = 2C_{1}\int_{\theta = 0}^{\pi}\sin\theta e^{\frac{1}{\lambda_{1}}\cos\theta\cos\phi}\sinh\left(\frac{1}{\lambda_{1}}\sin\theta\sin\phi\right)h_{0}(\cos\theta)d\theta,
\end{align*}  
where $C_1$ is the normalisation constant of the $\theta \rightarrow e^{\cos\theta/\lambda_1}$.
\item We simplify the last terms of the equilibrium equation (\ref{Eq:X_0_explicit}):
\begin{align*}
&(\text{\bf Id} -\bOmega_{0}\otimes\bOmega_{0})\int_{\bomega\in\S^{1}}(\bomega\otimes\bomega)M_{\lambda_1,\bOmega_{1}}h_{0}(\bomega\cdot\bOmega_{0})d\bomega \bOmega_{0} =\\ 
&\quad =(\text{\bf Id} -\bOmega_{0}\otimes\bOmega_{0})\left[\int_{\theta = -\pi}^{\pi}(\cos\theta)^2 C_{1}e^{\frac{1}{\lambda_{1}}\cos(\theta - \phi)} h_{0}(\cos\theta)d\theta\ (\bOmega_{0}\otimes\bOmega_{0})\bOmega_{0}\right.\\
&\quad +\int_{\theta = -\pi}^{\pi}\sin\theta\cos\theta C_{1}e^{\frac{1}{\lambda_{1}}\cos(\theta - \phi)} h_{0}(\cos\theta)d\theta \left[(\bOmega_{0}^{\perp}\otimes\bOmega_{0}) + (\bOmega_{0}\otimes\bOmega_{0}^{\perp})\right]\bOmega_{0}\\
&\quad \left.+\int_{\theta = -\pi}^{\pi}(\sin\theta)^2 C_{1}e^{\frac{1}{\lambda_{1}}\cos(\theta - \phi)} h_{0}(\cos\theta)d\theta\ (\bOmega_{0}^{\perp}\otimes\bOmega_{0}^{\perp})\bOmega_{0}\right]\\
&\quad = \int_{\theta = -\pi}^{\pi}\sin\theta\cos\theta C_{1}e^{\frac{1}{\lambda_{1}}\cos(\theta - \phi)} h_{0}(\cos\theta)d\theta\ \bOmega_{0}^{\perp}\\
&\quad = 2C_{1}\int_{\theta = 0}^{\pi}\sin\theta\cos\theta e^{\frac{1}{\lambda_{1}}\cos\theta\cos\phi}\sinh\left(\frac{1}{\lambda_{1}}\sin\theta\sin\phi\right)h_{0}(\cos\theta)d\theta\ \bOmega_{0}^{\perp}.
\end{align*} 
\end{enumerate}
So, we obtain from~(\ref{Eq:X_0_explicit}):
\begin{align*}
& 2C_1\int_{\theta = 0}^{\pi}\left(\frac{\rho_{1}}{\tau_{1}} + \frac{\alpha\rho_{0}\rho_{1}}{2\tau_{1}}(1 +c_{0}\cos\theta)\right)\sin\theta e^{\frac{1}{\lambda_{1}}\cos\theta\cos\phi}\sinh\left(\frac{1}{\lambda_{1}}\sin\theta\sin\phi\right) h_{0}(\cos\theta)d\theta\\
&\quad -\frac{\alpha c_{1}}{2\tau_{0}}\beta_{0}\rho_{0}\rho_{1} \sin\phi = 0.\nonumber
\label{Eq:momentum_balance_1_vers2}
\end{align*}
Since $(I_2)_{0}(\theta) = h_{0}(\cos\theta)\sin\theta$ is non positive on $(0,\pi)$ (application of the maximum principle for the elliptic equations \eqref{Eq:elliptic}), the two terms of the last equations have the same sign (the sign of $\sin\phi$) and thus the two equal zero. Therefore, the angle $\phi$ equals $0$ or $\pi$, which implies $\bOmega_0 = \bOmega_1$ or $\bOmega_0 = - \bOmega_1$.
\end{proof}

\section[Appendix: The linearised exchange operator]{Appendix: Proof of proposition~\ref{Prop:collisional_invariant} (The linearised exchange operator)}
\label{Appendix:CollisionalInvariant}

\begin{proof} We here provide a detailed proof of the computation of the exchange term: 
\begin{equation*}
(\text{\bf Id}-\bOmega_{0}\otimes\bOmega_{0})\left[(D\bS_0)_{(\rho_{0},\bOmega_0,\rho_{1},\bOmega_1)}(\tilde\rho_{0},\tilde\bOmega_0,\tilde\rho_{1},\tilde\bOmega_1)\right] - (\bOmega_0\cdot \bS_0(\rho_{0},\bOmega_0,\rho_{1},\bOmega_1))\tilde\bOmega_0
\end{equation*}
once the directions are at equilibria, i.e. $\bOmega_0 = \bOmega_1$ or $\bOmega_0 = -\bOmega_1$.
 
 Let us begin by developing the expressions of the two components, $\bX_0$ and $\bY_0$, of the linearised exchange operator $D\bS_0$ introduced in lemma~\ref{Prop:expansion_delta}:
\begin{align*}
\bX_0 =\ & \alpha\frac{\tilde\rho_0}{2}\rho_1\left[\int_{\bomega \in \S^1}\bomega M_{\lambda_1,\bOmega_1}  h_0d\bomega\right]\\
& +\left[\int_{\bomega \in \S^1}(\bomega\otimes\bomega)M_{\lambda_1,\bOmega_1}h_{0}d\bomega\right]\ (\alpha\frac{c_0\tilde\rho_0}{2}\rho_1\bOmega_0 + \alpha\frac{c_0\rho_0}{2}\rho_1\tilde\bOmega_0)\\
& + \left(1+\alpha\frac{\rho_0}{2}\right)\tilde\rho_1\left[\int_{\bomega \in \S^1}\bomega M_{\lambda_1,\bOmega_1}  h_0d\bomega\right] \\
&+\left[\int_{\bomega \in \S^1}(\bomega\otimes\bomega)M_{\lambda_1,\bOmega_1}h_{0}d\bomega\right]\ \left(\alpha\frac{c_0\rho_0}{2}\tilde\rho_1\bOmega_0 + (1+\alpha\frac{\rho_0}{2})\lambda_1^{-1}\rho_1\tilde\bOmega_1\right)\\
& +\left[\int_{\bomega \in \S^1}(\bomega\cdot\tilde\bOmega_1)(\bomega\otimes\bomega)M_{\lambda_1,\bOmega_1}h_{0}d\bomega\right]\ \alpha\frac{c_0\rho_0}{2}\lambda_1^{-1}\rho_1\bOmega_0\\
& + (1+\alpha\frac{\rho_0}{2})\rho_1\left[\int_{\bomega \in \S^1}(\bomega\otimes\bomega)M_{\lambda_1,\bOmega_1}h_{0}'d\bomega\right]\tilde\bOmega_0\\
& + \alpha\frac{c_0\rho_0}{2}\rho_1\left[\int_{\bomega \in \S^1}(\bomega\cdot\tilde\bOmega_0)(\bomega\otimes\bomega)M_{\lambda_1,\bOmega_1}h_{0}'d\bomega\right]\ \bOmega_0,\\[0.2cm]
\bY_0 =\ & \alpha\frac{\tilde\rho_1}{2}\rho_0\left[\int_{\bomega \in \S^1}\bomega M_{\lambda_0,\bOmega_0}  h_0d\bomega\right]\\
 & +\left[\int_{\bomega \in \S^1}(\bomega\otimes\bomega)M_{\lambda_0,\bOmega_0}h_{0}d\bomega\right]\ (\alpha\frac{c_1\tilde\rho_1}{2}\rho_0\bOmega_1 + \alpha\frac{c_1\rho_1}{2}\rho_0\tilde\bOmega_1)\\
& + \left(1+\alpha\frac{\rho_1}{2}\right)\tilde\rho_0\left[\int_{\bomega \in \S^1}\bomega M_{\lambda_0,\bOmega_0}  h_0d\bomega\right] \\
&+\left[\int_{\bomega \in \S^1}(\bomega\otimes\bomega)M_{\lambda_0,\bOmega_0}h_{0}d\bomega\right]\ \left(\alpha\frac{c_1\rho_1}{2}\tilde\rho_0\bOmega_1 + (1+\alpha\frac{\rho_1}{2})\lambda_0^{-1}\rho_0\tilde\bOmega_0\right)\\
& +\left[\int_{\bomega \in \S^1}(\bomega\cdot\tilde\bOmega_0)(\bomega\otimes\bomega)M_{\lambda_0,\bOmega_0}h_{0}d\bomega\right]\ \alpha\frac{c_1\rho_1}{2}\lambda_0^{-1}\rho_0\bOmega_1\\
& + (1+\alpha\frac{\rho_1}{2})\rho_0\left[\int_{\bomega \in \S^1}(\bomega\otimes\bomega)M_{\lambda_0,\bOmega_0}h_{0}'d\bomega\right]\tilde\bOmega_0\\
& + \alpha\frac{c_1\rho_1}{2}\rho_0\left[\int_{\bomega \in \S^1}(\bomega\cdot\tilde\bOmega_0)(\bomega\otimes\bomega)M_{\lambda_0,\bOmega_0}h_{0}'d\bomega\right]\ \bOmega_1,
\end{align*}
where $h_0$, $h_0'$  (resp. $h_1$, $h_1'$) implicitly depend on $(\bomega\cdot\bOmega_0)$ (resp. $(\bomega\cdot\bOmega_1)$). These expressions can be simplified when computed at equilibria. Indeed, the following lemma gives some useful identities. 

\begin{lem} We have the following identities:
\begin{align*}
\bullet\quad &\int_{\bomega \in \S^1}\bomega M_{\lambda,\bOmega}  h d\bomega = \langle\cos\theta h \rangle_{M_{\lambda}}\bOmega,\quad \int_{\bomega \in \S^1}\bomega M_{\lambda,-\bOmega}  h d\bomega = -\langle\cos\theta h^-\rangle_{M_{\lambda}}\bOmega,\\
\bullet\quad &\int_{\bomega \in \S^1}(\bomega\otimes\bomega)M_{\lambda,\bOmega}h d\bomega = \langle\cos^2\theta h\rangle_{M_{\lambda}}\bOmega\otimes\bOmega + \langle\sin^2\theta h\rangle_{M_{\lambda}}\bOmega^{\perp}\otimes\bOmega^{\perp},\\
&\int_{\bomega \in \S^1}(\bomega\otimes\bomega)M_{\lambda,-\bOmega}h d\bomega = \langle\cos^2\theta h^-\rangle_{M_{\lambda}}\bOmega\otimes\bOmega + \langle\sin^2\theta h^-\rangle_{M_{\lambda}}\bOmega^{\perp}\otimes\bOmega^{\perp},\\
\bullet\quad &\int_{\bomega \in \S^1}(\bomega\cdot\tilde\bOmega)(\bomega\otimes\bomega)M_{\lambda,\bOmega}h d\bomega = (\tilde\bOmega\cdot\bOmega^{\perp})\langle\sin^2\theta\cos\theta h\rangle_{M_{\lambda}}\left(\bOmega\otimes\bOmega^{\perp} + \bOmega^{\perp}\otimes\bOmega\right),\\
&\int_{\bomega \in \S^1}(\bomega\cdot\tilde\bOmega)(\bomega\otimes\bomega)M_{\lambda,-\bOmega}h d\bomega = -(\tilde\bOmega\cdot\bOmega^{\perp})\langle\sin^2\theta\cos\theta h^-\rangle_{M_{\lambda}}\left(\bOmega\otimes\bOmega^{\perp} + \bOmega^{\perp}\otimes\bOmega\right).
\end{align*}
where $h$ is a function depending on $\bomega\cdot\bOmega$ and $h^- : \bomega\cdot\bOmega \rightarrow h(-\bomega\cdot\bOmega)$ is the symmetric function of $h$. 
\end{lem}
\noindent Therefore, at equilibria $\bOmega_0 = \bOmega_1$, the two components of the linearised exchange operator, $\bX_0$ and $\bY_0$, simplified into:
\begin{align*}
\bX_0 =\ & \tilde\rho_0\left(\frac{\alpha}{2}\rho_1\langle\cos\theta h_0\rangle_{M_{\lambda_1}} + \frac{\alpha c_0}{2}\rho_1\langle\cos^2\theta h_0\rangle_{M_{\lambda_1}}\right)\bOmega_0\\
& + \tilde\rho_1\left((1+\alpha\frac{\rho_0}{2})\langle\cos\theta h_0\rangle_{M_{\lambda_1}} + \alpha\frac{c_0\rho_0}{2}\langle\cos^2\theta h_0\rangle_{M_{\lambda_1}}\right)\bOmega_0\\
&+\left(\alpha \frac{c_0\rho_0}{2}\rho_1\langle\sin^2\theta h_0\rangle_{M_{\lambda_1}} 
 + (1 +  \alpha\frac{\rho_0}{2})\rho_1\langle\sin^2\theta h_0'\rangle_{M_{\lambda_1}}\right.\\
&\hspace{1cm} \left.  + \alpha\frac{\rho_0 c_0}{2}\rho_1\langle\sin^2\theta\cos\theta h_0'\rangle_{M_{\lambda_1}} \right)\tilde\bOmega_0\\
&+ \left((1+\alpha\frac{\rho_0}{2})\lambda_1^{-1}\rho_1\langle\sin^2\theta h_0\rangle_{M_{\lambda_1}}
+ \alpha\frac{c_0\rho_0}{2}\rho_1\lambda_1^{-1}\langle\sin^2\theta\cos\theta h_0\rangle_{M_{\lambda_1}}\right)\tilde\bOmega_1,\\[0.2cm]
\bY_0 \ =& \tilde\rho_0\left((1+\alpha\frac{\rho_1}{2})\langle\cos\theta h_0\rangle_{M_{\lambda_0}} + \alpha\frac{c_1\rho_1}{2}\langle\cos^2\theta h_0\rangle_{M_{\lambda_0}}\right)\bOmega_0\\
&+ \tilde\rho_1\left(\frac{\alpha}{2}\rho_0\langle\cos\theta h_0\rangle_{M_{\lambda_0}} + \frac{\alpha c_1}{2}\rho_0\langle\cos^2\theta h_0\rangle_{M_{\lambda_0}}\right)\bOmega_0\\
&+\left((1+\alpha\frac{\rho_1}{2})\lambda_0^{-1}\rho_0\langle\sin^2\theta h_0\rangle_{M_{\lambda_0}}
 + \alpha\frac{c_1\rho_1}{2}\rho_0\lambda_0^{-1}\langle\sin^2\theta\cos\theta h_0\rangle_{M_{\lambda_0}}\right.\\
&\hspace{1cm}\left. + (1 + \alpha\frac{\rho_1}{2})\rho_0\langle\sin^2\theta h_0'\rangle_{M_{\lambda_0}}
+ \alpha\frac{c_1\rho_1}{2}\rho_0\langle\sin^2\theta\cos\theta h_0'\rangle_{M_{\lambda_0}}\right)\tilde\bOmega_0\\ 
&+\left(\alpha \frac{c_1\rho_1}{2}\rho_0\langle\sin^2\theta h_0\rangle_{M_{\lambda_0}}\right)\tilde\bOmega_1.
\end{align*}
Similarly, at equilibria $\bOmega_0 = \bOmega_1$, the momentum exchange operator $\bS_0$ can be expressed as follows: 
\begin{align*}
\bS_0(\rho_0,\Omega_0,\rho_1,\Omega_1) =\ & \tau_1\left((1+\alpha\frac{\rho_0}{2})\rho_1\langle\cos\theta h_0\rangle_{M_{\lambda_1}}  + \alpha\frac{c_0\rho_0}{2}\rho_1\langle\cos^2\theta h_0\rangle_{M_{\lambda_1}}\right)\bOmega_0\\
& - \tau_0\left((1+\alpha\frac{\rho_1}{2})\rho_0\langle\cos\theta h_0\rangle_{M_{\lambda_0}} + \alpha\frac{c_1\rho_1}{2}\rho_0\langle\cos^2\theta h_0\rangle_{M_{\lambda_0}}\right)\bOmega_0.
\end{align*}

\noindent Assuming $\bOmega_0 = \bOmega_1$, we thus obtain:
\begin{align*}
&(\text{\bf Id}-\bOmega_{0}\otimes\bOmega_{0})\left[(D\bS_0)_{(\rho_{0},\bOmega_0,\rho_{1},\bOmega_1)}(\tilde\rho_{0},\tilde\bOmega_0,\tilde\rho_{1},\tilde\bOmega_1)\right] - (\bOmega_0\cdot \bS_0(\rho_{0},\bOmega_0,\rho_{1},\bOmega_1))\tilde\bOmega_0\\
&\hspace{0.2cm}= \tau_1(1+\alpha\frac{\rho_0}{2})\rho_1\left[\lambda_1^{-1}\langle\sin^2\theta h_0\rangle_{M_{\lambda_1}}\tilde\bOmega_1 + \langle\sin^2\theta h_0'\rangle_{M_{\lambda_1}}\tilde\bOmega_{0} - \langle\cos\theta h_0\rangle_{M_{\lambda_1}}\tilde\bOmega_{0}\right]\\
&\hspace{0.5cm} +\tau_1\alpha \frac{c_0\rho_0}{2}\rho_1\left[\langle\sin^2\theta h_0\rangle_{M_{\lambda_1}}\tilde\bOmega_0 + \langle\sin^2\theta\cos\theta h_0'\rangle_{M_{\lambda_1}}\tilde\bOmega_0\right.\\
&\hspace{3cm}\left. + \lambda_1^{-1}\langle\sin^2\theta\cos\theta h_0\rangle_{M_{\lambda_1}}\tilde\bOmega_1 - \langle\cos^2\theta h_0\rangle_{M_{\lambda_1}}\tilde\bOmega_0\right]\\
&\hspace{0.5cm} - \tau_0(1+\alpha\frac{\rho_1}{2})\rho_0\left[\lambda_0^{-1}\langle\sin^2\theta h_0\rangle_{M_{\lambda_0}}\tilde\bOmega_0 + \langle\sin^2\theta h_0'\rangle_{M_{\lambda_0}}\tilde\bOmega_{0} - \langle\cos\theta h_0\rangle_{M_{\lambda_0}}\tilde\bOmega_{0}\right]\\
&\hspace{0.5cm} - \tau_0\alpha \frac{c_1\rho_1}{2}\rho_0\left[\langle\sin^2\theta h_0\rangle_{M_{\lambda_0}}\tilde\bOmega_1 + \langle\sin^2\theta\cos\theta h_0'\rangle_{M_{\lambda_0}}\tilde\bOmega_0\right.\\
&\hspace{3cm}\left. + \lambda_0^{-1}\langle\sin^2\theta\cos\theta h_0\rangle_{M_{\lambda_0}}\tilde\bOmega_0 - \langle\cos^2\theta h_0\rangle_{M_{\lambda_0}}\tilde\bOmega_0\right],\\
&\hspace{0.2cm} = A_{0}(\rho_0,\rho_1)(\tilde\bOmega_1 - \tilde\bOmega_0),
\end{align*}
with
 \begin{align*}
 A_{0}(\rho_0,\rho_1) =\ & \tau_1(1+\alpha\frac{\rho_0}{2})\rho_1\lambda_1^{-1}\langle\sin^2\theta h_0\rangle_{M_{\lambda_1}}\\ 
 &+ \tau_1\alpha \frac{c_0\rho_0}{2}\rho_1\lambda_1^{-1}\langle\sin^2\theta\cos\theta h_0\rangle_{M_{\lambda_1}} - \tau_0\alpha \frac{c_1\rho_1}{2}\rho_0\langle\sin^2\theta h_0\rangle_{M_{\lambda_0}},
\end{align*}
where we used the identities:
\begin{align*}
&\lambda_0^{-1}\langle\sin^2\theta h_0\rangle_{M_{\lambda_0}} = \langle\cos\theta h_0\rangle_{M_{\lambda_0}} - \langle\sin^2\theta h_0'\rangle_{M_{\lambda_0}},\\
&\lambda_0^{-1}\langle\sin^2\theta\cos\theta h_0\rangle_{M_{\lambda_0}} = \langle\cos^2\theta h_0\rangle_{M_{\lambda_0}} - \langle\sin^2\theta h_0\rangle_{M_{\lambda_0}} - \langle\sin^2\theta\cos\theta h_0'\rangle_{M_{\lambda_0}}.
\end{align*}

Similar computations shows that at equilibria $\bOmega_0 = -\bOmega_1$,
\begin{align*}
&(\text{\bf Id}-\bOmega_{0}\otimes\bOmega_{0})\left[(D\bS_0)_{(\rho_{0},\bOmega_0,\rho_{1},\bOmega_1)}(\tilde\rho_{0},\tilde\bOmega_0,\tilde\rho_{1},\tilde\bOmega_1)\right] - (\bOmega_0\cdot \bS_0(\rho_{0},\bOmega_0,\rho_{1},\bOmega_1))\tilde\bOmega_0\\
&\hspace{0.2cm} = B_{0}(\rho_0,\rho_1)(\tilde\bOmega_1 + \tilde\bOmega_0),
\end{align*}
with
 \begin{align*}
B_{0}(\rho_0,\rho_1) =\ & \tau_1(1+\alpha\frac{\rho_0}{2})\rho_1\lambda_1^{-1}\langle\sin^2\theta h_0^-\rangle_{M_{\lambda_1}}\\ 
 &- \tau_1\alpha \frac{c_0\rho_0}{2}\rho_1\lambda_1^{-1}\langle\sin^2\theta\cos\theta h_0^-\rangle_{M_{\lambda_1}} - \tau_0\alpha \frac{c_1\rho_1}{2}\rho_0\langle\sin^2\theta h_0\rangle_{M_{\lambda_0}}.
\end{align*}
\end{proof}

\paragraph{Acknowledgment} 
The author would like to express his gratitude to Pierre Degond and Francesco Ginelli for their fruitful suggestions. He wishes also to thank Richard Bon, Hugues Chat\'e, Marie-H\'el\`ene Pillot and Guy Th\'eraulaz for stimulating discussions. This work has been supported by the ``Agence Nationale de la Recherche'', in the frame of the contract PANURGE 07-BLAN-0208.

\bibliography{Vicsek_diphasic}

\begin{thebibliography}{10}

\bibitem{2003_AldanaHuepe}
M.~Aldana and C.~Huepe.
\newblock {Phase Transitions in Self-Driven Many-Particle Systems and Related
  Non-Equilibrium Models: A Network Approach}.
\newblock {\em J. Stat. Phys.}, 112(1):135--153, 2003.

\bibitem{1982_Aoki}
I.~Aoki.
\newblock A simulation study on the schooling mechanism in fish.
\newblock {\em Bull. Jpn. Soc. Sci. Fisher}, 48:1081--1088, 1982.

\bibitem{2008_Ballerini}
M.~Ballerini, N.~Cabibbo, R.~Candelier, A.~Cavagna, E.~Cisbani, I.~Giardina,
  V.~Lecomte, A.~Orlandi, G.~Parisi, A.~Procaccini, M.~Viale, and
  V.~Zdravkovic.
\newblock Interaction ruling animal collective behavior depends on topological
  rather than metric distance: Evidence from a field study.
\newblock {\em Proc. National Academy of Sciences}, 105(4):1232--1237, 2008.

\bibitem{2010_Bellomo_Multiscale}
N.~Bellomo, A.~Bellouquid, J.~Nieto, and J.~Soler.
\newblock {Multiscale biological tissue models and flux-limited chemotaxis for
  multicellular growing systems}.
\newblock {\em Math. Models Methods Appl. Sci}, 20:1179--1207, 2010.

\bibitem{2008_BellomoDelitala_cells}
N.~Bellomo and M.~Delitala.
\newblock From the mathematical kinetic, and stochastic game theory to
  modelling mutations, onset, progression and immune competition of cancer
  cells.
\newblock {\em Phys. Life. Rev.}, 5(4):183--206, 2008.

\bibitem{2011_BellomoDogbe_traffic}
N.~Bellomo and C.~Dogb\'e.
\newblock On the modeling of traffic and crowds: A survey of models,
  speculations, and perspectives,.
\newblock {\em SIAM Rev.}, 53(3):409--463, 2011.

\bibitem{2005_BellouDellitala}
A.~Bellouquid and M.~Delitala.
\newblock Mathematical methods and tools of kinetic theory towards modelling
  complex biological systems.
\newblock {\em Math. Models Methods Appl. Sci}, 15(11):1639--1666, 2005.

\bibitem{2006_BoltzSelfPropel_BertinGregoire}
E.~Bertin, M.~Droz, and G.~Gr{\'e}goire.
\newblock Boltzmann and hydrodynamic description for self-propelled particles.
\newblock {\em Phys. Rev. E.}, 74:022101, 2006.

\bibitem{2010_BolleyCanCar_StocMFlimit}
F.~Bolley, J.A. Ca{\~n}izo, and J.A. Carrillo.
\newblock {Stochastic Mean-Field Limit: Non-Lipschitz Forces \& Swarming}.
\newblock {\em Math. Models Methods Appl. Sci.}, 21(11):2179--2210, 2011.

\bibitem{2011_BolleyCanCar_Vicsek}
F.~Bolley, J.A. Ca{\~n}izo, and J.A. Carrillo.
\newblock {Mean-field limit for the stochastic Vicsek model}.
\newblock {\em Appl. Math. Lett.}, 25(3):339--343, 2012.

\bibitem{2010_CarFornRosTosc_AsymptFlock}
J.~A. Carrillo, M.~Fornasier, J.~Rosado, and G.~Toscani.
\newblock Asymptotic flocking dynamics for the kinetic cucker-smale model.
\newblock {\em SIAM J. Math. Anal.}, 42:218--236, 2010.

\bibitem{2009_Carillo_DbleMilling}
J.A. Carrillo, M.R. D'orsogna, and V.~Panferov.
\newblock Double milling in self-propelled swarms from kinetic theory.
\newblock {\em Kinet. Relat. Models}, 2:363--378, 2009.

\bibitem{2010_CarilloForn_Review}
J.A. Carrillo, M.~Fornasier, G.~Toscani, and F.~Vecil.
\newblock Particle, kinetic, and hydrodynamic models of swarming.
\newblock {\em Mathematical Modeling of Collective Behavior in Socio-Economic
  and Life Sciences}, 297--336, 2010.

\bibitem{2010_CarilloKlar}
J.A. Carrillo, A.~Klar, S.~Martin, and S.~Tiwari.
\newblock Self-propelled interacting particle systems with roosting force.
\newblock {\em Math. Models Methods Appl. Sci}, 20:1533--1552, 2010.

\bibitem{2010_Cavagna_EmpData}
A.~Cavagna, A.~Cimarelli, I.~Giardina, G.~Parisi, R.~Santagati, F.~Stefanini,
  and R.~Tavarone.
\newblock From empirical data to inter-individual interactions: unveiling the
  rules of collective animal behavior.
\newblock {\em Math. Models Methods Appl. Sci}, 20:1491--1510, 2010.

\bibitem{2008_ChateGin_CollMot}
H.~Chat{\'e}, F.~Ginelli, G.~Gr{\'e}goire, and F.~Raynaud.
\newblock {Collective motion of self-propelled particles interacting without
  cohesion}.
\newblock {\em Phys. Rev. E}, 77(4):046113, 2008.

\bibitem{2007_DOrsognaBertozzi}
Y.~Chuang, M.R. D'Orsogna, D.~Marthaler, A.L. Bertozzi, and L.S. Chayes.
\newblock {State transitions and the continuum limit for a 2D interacting,
  self-propelled particle system}.
\newblock {\em Physica D}, 232(1):33--47, 2007.

\bibitem{2003_Vertebrates_CouzinKrause}
I.D. Couzin and J.~Krause.
\newblock {Self-organization and collective behavior in vertebrates}.
\newblock {\em Adv. Stud. Behav.}, 32(1), 2003.

\bibitem{2002_Couzin}
I.D. Couzin, J.~Krause, R.~James, G.D. Ruxton, and N.R. Franks.
\newblock Collective memory and spatial sorting in animal groups.
\newblock {\em J. Theoret. Biol.}, 218(1):1--11, 2002.

\bibitem{2007_EmergenceFlocks_CuckerSmale}
F.~Cucker and S.~Smale.
\newblock {Emergent behavior in flocks}.
\newblock {\em IEEE Trans. Automat. Control}, 52(5):852--862, 2007.

\bibitem{2000_CzirokVicsek}
A.~Czir{\'o}k and T.~Vicsek.
\newblock Collective behavior of interacting self-propelled particles.
\newblock {\em Physica A}, 281(1-4):17--29, 2000.

\bibitem{2008_ContinuumLimit_DM}
P.~Degond and S.~Motsch.
\newblock {Continuum limit of self-driven particles with orientation
  interaction}.
\newblock {\em Math. Models Methods Appl. Sci.}, 18:1193--1215, 2008.

\bibitem{2011_DegondMotsch_PTWMacro}
P.~Degond and S.~Motsch.
\newblock A macroscopic model for a system of swarming agents using curvature
  control.
\newblock {\em J. Stat. Phys.}, 143:685--714, 2011.

\bibitem{2010_DegondNavoret_Congestion}
P.~Degond, L.~Navoret, R.~Bon, and D.~Sanchez.
\newblock Congestion in a macroscopic model of self-driven particles modeling
  gregariousness.
\newblock {\em J. Stat. Phys.}, 138(1):85--125, 2010.

\bibitem{2007_DelitalaTosin_DiscreteKinetic}
M.~Delitala and A.~Tosin.
\newblock Mathematical modeling of vehicular traffic: a discrete kinetic theory
  approach.
\newblock {\em Math. Models Methods Appl. Sci.}, 17(6):901--932, 2007.

\bibitem{2010_Frouvelle_Vic}
A.~Frouvelle.
\newblock A continuum model for alignment of self-propelled particles with
  anisotropy and density-dependent parameters.
\newblock {\em Math. Models Methods Appl. Sci.}, to appear, 2012.

\bibitem{2011_FrouvLiu_PhaseTrans}
A.~Frouvelle and J.G. Liu.
\newblock {Dynamics in a kinetic model of oriented particles with phase
  transition}.
\newblock {\em SIAM J. Math. Anal.}, to appear, 2012.

\bibitem{1975_Gatignol}
R.~Gatignol.
\newblock {\em {Th\'eorie cin\'etique des gaz \`a r\'epartition discr\`ete de
  vitesses}}.
\newblock Lecture Notes in Physics 36, Springer Verlag, 1975.

\bibitem{2004_GregChate_Onset}
G.~Gr{\'e}goire and H.~Chat{\'e}.
\newblock Onset of collective and cohesive motion.
\newblock {\em Phys. Rev. Lett.}, 92(2):25702, 2004.

\bibitem{2009_ProofCuckerSmale_HaLiu}
S.Y. Ha and J.G. Liu.
\newblock {A simple proof of the Cucker-Smale flocking dynamics and mean-field
  limit}.
\newblock {\em Comm. Math. Sci.}, 7(2):297--325, 2009.

\bibitem{2008_Flocking_HaTadmor}
S.Y. Ha and E.~Tadmor.
\newblock {From particle to kinetic and hydrodynamic descriptions of flocking}.
\newblock {\em Kinet. Relat. Models}, 1(3):415--435, 2008.

\bibitem{2001_Helbing_RevTraffic}
D.~Helbing.
\newblock Traffic and related self-driven many-particle systems.
\newblock {\em Rev. Mod. Phys.}, 73(4):1067, 2001.

\bibitem{1975_Ishii}
M.~Ishii.
\newblock {\em Thermo-fluid dynamic theory of two-phase flow}.
\newblock Eyrolles, Paris, 1975.

\bibitem{1999_NonLocalModel_MogKesh}
A.~Mogilner and L.~Edelstein-Keshet.
\newblock {A non-local model for a swarm}.
\newblock {\em J. Math. Biol.}, 38(6):534--570, 1999.

\bibitem{2010_MotschLN_NumMacroVic}
S.~Motsch and L.~Navoret.
\newblock Numerical simulations of a non-conservative hyperbolic system with
  geometric constraints describing swarming behavior.
\newblock {\em Multiscale Model. Simul.}, 9(3):1253--1275, 2010.

\bibitem{2010_Pillot_CollMovInitStop}
M.H. Pillot and J.L. Deneubourg.
\newblock {Collective movements, initiation and stops: diversity of situations
  and law of parsimony}.
\newblock {\em Behav. Processes}, 84(3):657--661, 2010.

\bibitem{2011_Pillot_Scalable}
M.H. Pillot, J.~Gautrais, P.~Arrufat, I.D. Couzin, R.~Bon, and J.L. Deneubourg.
\newblock {Scalable Rules for Coherent Group Motion in a Gregarious
  Vertebrate}.
\newblock {\em PloS one}, 6(1):e14487, 2011.

\bibitem{2009_Pillot}
M.H. Pillot, J.~Gautrais, J.~Gouello, P.~Michelena, A.~Sibbald, and R.~Bon.
\newblock Moving together: incidental leaders and na\"ive followers.
\newblock {\em Behav. Processes}, 83(3):235--241, 2010.

\bibitem{spohn_book}
H.~Spohn.
\newblock {\em {Large Scale Dynamics of Interacting Particles}}.
\newblock Springer-Verlag New York, 1991.

\bibitem{2005_Stroock_MarkovProcess}
D.W. Stroock.
\newblock {\em An introduction to Markov processes}, volume 230.
\newblock Springer Verlag, 2005.

\bibitem{06_Nonlocal_TopazBertozzi}
C.M. Topaz, A.L. Bertozzi, and M.A. Lewis.
\newblock {A Nonlocal Continuum Model for Biological Aggregation}.
\newblock {\em Bull. Math. Biol.}, 68(7):1601--1623, 2006.

\bibitem{95_Vicsek_PhasTrans2d}
T.~Vicsek, A.~Czir{\'o}k, E.~Ben-Jacob, I.~Cohen, and O.~Shochet.
\newblock {Novel Type of Phase Transition in a System of Self-Driven
  Particles}.
\newblock {\em Phys. Rev. Lett.}, 75(6):1226--1229, 1995.

\end{thebibliography}
\bibliographystyle{plain}

\end{document}